\documentclass[twocolumn]{aastex7}
\usepackage{amsmath, mathtools, amsthm, amssymb}
\usepackage{enumitem}
\usepackage[toc,page]{appendix}
\hypersetup{colorlinks=true, allcolors=cyan}
\usepackage{soul}
\usepackage{graphicx}
\usepackage{appendix}
\graphicspath{ {./} }
\usepackage{xcolor}
\usepackage{hyperref}
\usepackage{nameref}

\usepackage[super]{nth}

\newcommand\soutm{\bgroup\markoverwith{\textcolor{black}{\rule[0.5ex]{2pt}{0.8pt}}}\ULon}

\newcommand{\msol}{\ensuremath{\rm{M_{\odot}}}}
\newcommand{\NH}{\textsc{NewHorizon}}
\newcommand{\NHII}{\textsc{NewHorizon2}}
\newcommand{\hcc}[1]{\ensuremath{#1\,\rm{H\,cm^{-3}}}}
\definecolor{Red}{rgb}{0.65,0.08,0.05}
\definecolor{Green}{rgb}{0.05,0.6,0.1}
\definecolor{azure}{rgb}{0.0, 0.5, 1.0}
\definecolor{Darkorange}{rgb}{0.7, 0.24, 0.03}

\newcommand\katdel{\bgroup\markoverwith{\textcolor{olive}{\rule[0.5ex]{2pt}{0.8pt}}}\ULon}

\newcommand\edel{\bgroup\markoverwith{\textcolor{Red}{\rule[0.5ex]{2pt}{0.8pt}}}\ULon}

\shorttitle{NH Missing Satellites and Starless subhalos}
\shortauthors{Jeon et al.}

\begin{document}

\title{Born to be Starless: Revisiting the Missing Satellite Problem}

\author{Seyoung Jeon}\email{syj3514@yonsei.ac.kr}
\affil{Department of Astronomy and Yonsei University Observatory, Yonsei University, 50 Yonsei-ro, Seodaemun-gu, Seoul 03722, Republic of Korea}

\author{Sukyoung K. Yi}\email{yi@yonsei.ac.kr}
\affil{Department of Astronomy and Yonsei University Observatory, Yonsei University, 50 Yonsei-ro, Seodaemun-gu, Seoul 03722, Republic of Korea}

\author{Emanuele Contini}\email{emanuele.contini82@gmail.com}
\affil{Department of Astronomy and Yonsei University Observatory, Yonsei University, 50 Yonsei-ro, Seodaemun-gu, Seoul 03722, Republic of Korea}

\author{Yohan Dubois}\email{dubois@iap.fr}
\affil{Institut d’ Astrophysique de Paris, Sorbonne Universités, et CNRS, UMP 7095, 98 bis bd Arago, 75014 Paris, France}

\author{San Han}\email{san.han@iap.fr}
\affil{Institut d’ Astrophysique de Paris, Sorbonne Universités, et CNRS, UMP 7095, 98 bis bd Arago, 75014 Paris, France}

\author{Katarina Kraljic}\email{kraljic@unistra.fr}
\affil{Observatoire astronomique de Strasbourg, Université de Strasbourg, CNRS, UMR 7550, Strasbourg 67000, France}

\author{Sebastien Peirani}\email{sebpeirani@gmail.com}
\affil{ILANCE, CNRS – University of Tokyo International Research Laboratory, Kashiwa, Chiba 277-8582, Japan}
\affil{Kavli IPMU (WPI), UTIAS, The University of Tokyo, Kashiwa, Chiba 277-8583, Japan}
\affil{Institut d’ Astrophysique de Paris, Sorbonne Universités, et CNRS, UMP 7095, 98 bis bd Arago, 75014 Paris, France}

\author{Christophe Pichon}\email{pichon@iap.fr}
\affil{Institut d’ Astrophysique de Paris, Sorbonne Universités, et CNRS, UMP 7095, 98 bis bd Arago, 75014 Paris, France}
\affil{Kyung Hee University, Dept. of Astronomy \& Space Science, Yongin-shi, Gyeonggi-do 17104, Republic of Korea}

\author{Jinsu Rhee}\email{jinsu.rhee@gmail.com}
\affil{Department of Astronomy and Yonsei University Observatory, Yonsei University, 50 Yonsei-ro, Seodaemun-gu, Seoul 03722, Republic of Korea}
\affil{Korea Astronomy and Space Science Institute, 776 Daedeokdae-ro, Yuseong-gu, Daejeon 34055, Republic of Korea}
\affil{Institut d’ Astrophysique de Paris, Sorbonne Universités, et CNRS, UMP 7095, 98 bis bd Arago, 75014 Paris, France}

%%%%%%%%%%%%%%%%%%%%%%%%%%%%%%%%%%%%%%%%%%%%%%%%%%%%%%%%%%%%%%%%%%%%%%%%%%%
%  _______  _______  _______  _______  ______    _______  _______  _______ 
% |   _   ||  _    ||       ||       ||    _ |  |   _   ||       ||       |
% |  |_|  || |_|   ||  _____||_     _||   | ||  |  |_|  ||       ||_     _|
% |       ||       || |_____   |   |  |   |_||_ |       ||       |  |   |  
% |       ||  _   | |_____  |  |   |  |    __  ||       ||      _|  |   |  
% |   _   || |_|   | _____| |  |   |  |   |  | ||   _   ||     |_   |   |  
% |__| |__||_______||_______|  |___|  |___|  |_||__| |__||_______|  |___|  
%%%%%%%%%%%%%%%%%%%%%%%%%%%%%%%%%%%%%%%%%%%%%%%%%%%%%%%%%%%%%%%%%%%%%%%%%%%
\begin{abstract}
The massive Local Group galaxies both host substantially fewer satellites than the subhalos expected from the cold dark matter paradigm, and the recent investigations have highlighted the interplay between baryons and dark matter. We investigate the processes that make subhalos starless, using high-resolution cosmological simulations. We found that the number of satellites around Milky Way analogs closely aligns with observations, which accords with recent studies. In our simulations, the majority of subhalos are devoid of stars, i.e., ``starless.'' We first examined supernova feedback and the environmental effects associated with subhalos' orbital motion as candidates of origin. However, neither seems to be the main driver. Supernova feedback causes a reduction of cold gas in ``starred'' subhalos, but its impact is not significant. In the case of starless subhalos, supernova feedback is irrelevant because most of them do not have in-situ star formation in the first place. The orbital motion in dense environments causes gas removal in all subhalos but is not enough to remove pre-existing stars. The key is found to be the effect of reionization instead. Starless subhalos are initially born in regions that are less efficient in accreting matter. This makes them lack sufficiently dense gas to self-shield from UV background heating, preventing their gas from cooling below the star formation threshold. This indicates that starless subhalos are not made but born.
\end{abstract}


%%%%%%%%%%%%%%%%%%%%%%%%%%%%%%%%%%%%%%%%%%%%%
%  ___   __    _  _______  ______    _______ 
% |   | |  |  | ||       ||    _ |  |       |
% |   | |   |_| ||_     _||   | ||  |   _   |
% |   | |       |  |   |  |   |_||_ |  | |  |
% |   | |  _    |  |   |  |    __  ||  |_|  |
% |   | | | |   |  |   |  |   |  | ||       |
% |___| |_|  |__|  |___|  |___|  |_||_______|
%%%%%%%%%%%%%%%%%%%%%%%%%%%%%%%%%%%%%%%%%%%%%
\section[]{Introduction}
\label{sec_intro}

% =================================================================
% > Bring up the missing satellite problem
Since the last decades, simulations based on the Lambda cold dark matter ($\Lambda$CDM, \citealt{Spergel2003, Planck2020}) cosmology have successfully reproduced large-scale structures of the Universe, enhancing our understanding of structure formation and galaxy evolution \citep{Dubois2014, Vogelsberger2014, Schaye2015, Mccarthy2017, Pillepich2018, dave2019, Dubois2021, Pakmor2023}.
However, this model has historically faced several small-scale tensions: e.g., the missing satellite problem \citep{Moore1999, Klypin1999, Boylan-Kolchin2011}, the cusp-core problem \citep{Flores1994, Moore1994}, and the plane of satellites \citep{Lynden-Bell1976, Pawlowski2012}.
The missing satellite problem traditionally refers to the finding that the number of predicted dark matter (DM) subhalos in DM-only simulations are far larger than that of the observed satellite galaxies around the Milky Way \citep{Klypin1999} and similar galaxies \citep{Tanaka2018}.
Thanks to the advanced observational instruments and techniques, the number of observed satellite galaxies has been increasing \citep{Grcevich2009, McConnachie2012, Mao2021, Geha2024}.
This progress, combined with improved completeness corrections and forward modeling techniques \citep{Kim2018, Homma2019, Nadler2020}, has significantly alleviated much of the original discrepancy for luminous dwarf galaxies; however, the discrepancy still remains, depending on the modeling.
Many studies have attempted to resolve this issue by adopting different approaches (see, e.g., \citealt{Read2019} and references therein).

% =================================================================
% > Suggested two solutions: dark matter solution and baryonic solution
One deals with alternative DM models by directly reducing the number of DM halos or modifying internal halo properties.
Warm DM (WDM) and fuzzy DM (FDM) suppress the formation of low-mass halos through free-streaming or quantum pressure, respectively, leading to a reduced number of small-scale substructures \citep{Hu2000, Gotz2002, Polisensky2011, Lovell2012, Marsh2016}.
For instance, \citet{Polisensky2011} used N-body simulations to show that WDM can reduce the number of low-mass halos and suggested the lower limit WDM particle mass of $2.3\,\rm keV$ to match satellite counts.
However, their analysis excluded baryonic effects, and the suggested mass is now disfavored by Lyman-$\alpha$ forest data \citep{Schneider2014}.
More recently, \citet{Nadler2021} suggested the forward-modeling framework with $m_{\rm WDM}>6.5\,\rm keV$ can match the fitting results from DES and Pan-STARRS1, and \citet{Irsic2024} updated the constraints of $m_{\rm WDM}>5.7\,\rm keV$ from quasar spectra.
Moreover, \citet{Garcia-Gallego2025} showed that the extreme WDM of $m_{\rm WDM}=1\,\rm keV$ can be viable with the upper limit of WDM fraction of 0.16.
Self-interacting DM (SIDM), on the other hand, does not significantly alter halo abundance but transforms the internal structure of halos with flattening central density profiles and adjusting maximum circular velocities, which gives better comparison with observed dynamical mass and profiles \citep{Spergel2000, vandenAarssen2012, Vogelsberger2016}.

Alternatively, the effect of the baryonic components could result in a reduced number of satellite galaxies without modifying the nature of $\Lambda$CDM.
The proper inclusion of baryons influences the DM distribution \citep{Zolotov2012, Wetzel2016, Garrison-Kimmel2019} and star formation rates in dwarf galaxies through feedback processes, such as from supernovae \citep{Brooks2013, Geen2013, Hazenfratz2024}, active galactic nuclei \citep[AGN; ][]{Dashyan2018, Arjona2024}, and ionizing UV radiation \citep{Sawala2016, Forbes2016, Emerick2019}.
Numerous studies explored baryonic solutions in simulations and other simplistic models and have succeeded in alleviating the classical missing satellite problem, but they often lack resolutions sufficient to resolve the problem entirely in a fully cosmological framework.

Moreover, these studies have not fully addressed the mechanisms responsible for the widespread transformation of subhalos into starless systems.
Reionization, in particular, has long been recognized as a key mechanism in suppressing galaxy formation in low-mass halos.
Early analytic and semi-analytic works \citep{Efstathiou1992, Bullock2000, Benson2002, Somerville2002} demonstrated that heating of the intergalactic medium after reionization raises the Jeans mass, preventing gas accretion into halos below a critical virial temperature (e.g., $\sim2\times10^4\,$K).
A comprehensive review by \citet{Somerville2015} further highlights how this mechanism naturally limits star formation in halos with $M_{\rm vir} \lesssim 10^9\,\msol$.
However, the interplay of reionization, internal feedback, and environmental effects, especially within Local Group-like systems, still needs to be quantified using cosmological simulations with sufficiently high resolution.
For example, the timing of a halo’s mass assembly, which is highly affected by cosmological environments, is known to strongly influence its ability to retain gas and form stars, not just its mass at $z=0$ \citep{Okamoto2008}.
Understanding this process is crucial for explaining the population of subhalos, and it constitutes the central focus of this work.
We hereby focus on the baryonic effects without touching the cosmological framework.

% =================================================================
% > Needs for simulation
From the observational perspective, it is difficult to address this issue empirically because starless dark halos are virtually unobservable.
Furthermore, analytical studies have shown that the gas content of halos is highly sensitive to the timing of reionization \citep{Noh2014}.
These motivate the need for high-resolution simulations that can track gas retention and star formation across diverse environments.
However, as modern hydrodynamical simulations require significantly more computational resources than DM-only simulations, achieving a balance between large cosmological volumes and the high resolution needed to resolve faint dwarf satellites remains challenging.
This compromise often limits our ability to simultaneously capture both the development of galaxies in the cosmological context and the detailed structure of dwarf galaxies.
This is why previous simulation-based studies have often focused on isolated dwarf galaxies or a limited number of Milky Way analogs, restricting their ability to explore environmental diversity and statistically significant subhalo populations \citep{Munshi2017, Applebaum2021}.

% =================================================================
% > Introducing the NH and NH2
To address this limitation, we employ high-resolution cosmological simulations, \NH\ \citep{Dubois2021} and \NHII\ \citep{Yi2024} whose details are given in Section~\ref{sec_method_sim}.
High resolution is essential to resolve dense star-forming regions, which is required for studying the formation of faint objects.
% =================================================================
% > Merit of our simulations
Our simulations successfully reproduce the morphology of observed spiral galaxies with thin disk structures and small star-forming clumps \citep{Yi2024}.
Also, these cosmological zoom-in simulations cover a spherical volume with a diameter of 20 Mpc, which is an exceptional scale in simulations under 100 pc resolution.
This ensures a relatively larger sample size evolving in the cosmological context compared to idealized simulations targeting individual galaxies.
Thus, we can effectively capture the formation and evolution of small-scale dwarf galaxies.
Furthermore, these simulations incorporate a comprehensive treatment of baryon physics, which is crucial for our analysis (see details in \citealt{Dubois2021}).

% =================================================================
% > Contents
The remainder of the paper is structured as follows.
In Section~\ref{sec_methods}, we provide a detailed description of our simulations, including the methods used for halo identification and subhalo classification, incorporating information from the merger trees.
Section~\ref{sec_results} presents our main findings, which are further discussed in Section~\ref{sec_discussion}, where we compare our results with previous studies and explore their broader implications.
Finally, in Section~\ref{sec_conclusion}, we summarize our key conclusions and outline potential directions for future research.
Throughout this work, stellar masses have been computed assuming a Chabrier \citep{Chabrier2003} initial mass function, and all masses are h-corrected.

%%%%%%%%%%%%%%%%%%%%%%%%%%%%%%%%%%%%%%%%%%%%%%%%%%%%%%%%%%%%%%%%%%%%%%%%%%%
%  __   __  _______  _______  __   __  _______  ______   _______  
% |  |_|  ||       ||       ||  | |  ||       ||      | |       |
% |       ||    ___||_     _||  |_|  ||   _   ||  _    ||  _____|
% |       ||   |___   |   |  |       ||  | |  || | |   || |_____ 
% |       ||    ___|  |   |  |       ||  |_|  || |_|   ||_____  |
% | ||_|| ||   |___   |   |  |   _   ||       ||       | _____| |
% |_|   |_||_______|  |___|  |__| |__||_______||______| |_______|
%%%%%%%%%%%%%%%%%%%%%%%%%%%%%%%%%%%%%%%%%%%%%%%%%%%%%%%%%%%%%%%%%%%%%%%%%%%
\section{Methods}
\label{sec_methods}

% =================================================================
% > Opening methods section
In the following, we provide a more detailed description of our set of simulations, outlining their key characteristics.
Given the critical role of halo and subhalo identification in our analysis, we also present a comprehensive explanation of the method used to detect them.
Additionally, we describe the approach adopted to establish a connection between halos, subhalos, and their associated galaxies (if any), highlighting any matching criteria applied in our study.

%%%%%%%%%%%%%%%%%%%%%%%%%%%%%%%%%%%%%%%%%%%%%%%%%%%%%%%%%%%%%%%%%%%%%%%%%%%
%%%%% 	[Numerical Simulation]
%%%%%%%%%%%%%%%%%%%%%%%%%%%%%%%%%%%%%%%%%%%%%%%%%%%%%%%%%%%%%%%%%%%%%%%%%%%
\subsection{Numerical simulations}\label{sec_method_sim}
% =================================================================
% > General Introduction
Here we describe detailed information about our simulations: \NH\ \citep{Dubois2021} and \NHII\ \citep{Yi2024}.
\NH\ is a follow-up to the cosmological hydrodynamical simulation Horizon-AGN \citep{Dubois2014}, which was run using the adaptive mesh refinement code, RAMSES \citep{Teyssier2002}.
Horizon-AGN provided numerous galaxy samples over a large volume of (140 Mpc)$^3$, but its kpc-scale resolution was insufficient to resolve faint dwarf galaxies.
To achieve higher resolution, \NH\ targets a smaller spherical volume with $20\,\rm Mpc$ diameter using the zoom-in technique.
The best spatial resolution is $\rm34\,pc$, and the stellar (DM) mass resolution is $\rm10^4\,\msol$ ($\rm10^6\,\msol$).
With this high resolution and modern astrophysical prescriptions, \NH\ successfully resolves small-scale galactic features such as thin discs, star-forming clumps, and dwarf galaxies \citep{Dubois2021, Martin2022, Yi2024, Rhee2024}.
Although the detailed subgrid astrophysics can be found in \citet{Dubois2021}, we want to introduce key physics very briefly.
For the star formation, each leaf cell should exceed the gas density threshold 5 (NH2) or \hcc{10} (NH).
Then, stars are finally formed if the cell satisfies the gravo-turbulent condition (i.e., local gravity and thermal/turbulent pressure; \citealt{Federrath2012}).
Our simulations follow the mechanical feedback scheme of \citet{Kimm2015} for Type II supernovae, which considers the energy release and momentum transfer from massive stars.
We also have the radiative heating source from the uniform UV background following the \citet{Haardt1996} model that is turned on at $z=10$.

% =================================================================
% > What's new in NH2
To enhance the statistics in our main work, we also adopt \NHII\ simulation that is a twin simulation of \NH\ with the same initial condition but with poorer spatial (68 pc) and stellar mass  ($2\times10^4\,\msol$) resolutions.
In addition to its lower resolution, \NHII\ has several minor differences.
Firstly, \NHII\ has a 50\% boosted stellar feedback efficiency than \NH, expecting that this improves the agreement of star formation in low-mass halos.
Secondly, \NHII\ employs a more sophisticated method for selecting the zoom-in region.
Like \NH, if we choose a simple spherical zoom-in region, the outer boundary of the zoom-in region suffers from artificial boundary effects caused by low-resolution DM particles (i.e., so-called ``contamination'').
Since even a few low-resolution particles could artificially distort the gravitational potential in halos, we had to discard many samples due to contamination in \NH.
Instead of a simple spherical region, \NHII\ selects the zoom-in region using the convex hull of the DM particles traced back to the initial condition.
This method significantly reduces contamination, which is especially crucial for small-scale structures.
Lastly, \NHII\ tracks the evolution of nine chemical elements (H, C, N, O, Mg, Si, S, Fe, and D).
Although chemical evolution is not the main topic of this work, this information has been used in galactic studies, such as the decomposition of the thin and thick discs \citep{Yi2024}.
Despite sacrificing resolution, many advanced techniques in \NHII\ require significantly more computational resources.
We overcome this issue by using the OpenMP and MPI hybrid parallelization version of the code, RAMSES-yOMP \citep{Han2024}.

% =================================================================
% > Good output time cadence
We also want to emphasize that we have a dense output cadence of $\sim1,000$ snapshots with an average time interval of 15 Myr for both simulations.
This enables us to capture short timescale events (e.g., the reaction of surrounding gas to stellar and AGN feedback) and detailed evolutionary tracks even in violent interactions with neighboring objects.

%=================================================================%
%          [Figure 01] (One Column)
%=================================================================%
\begin{figure}[tb!]
    % \centering
    \includegraphics[width=0.45\textwidth]{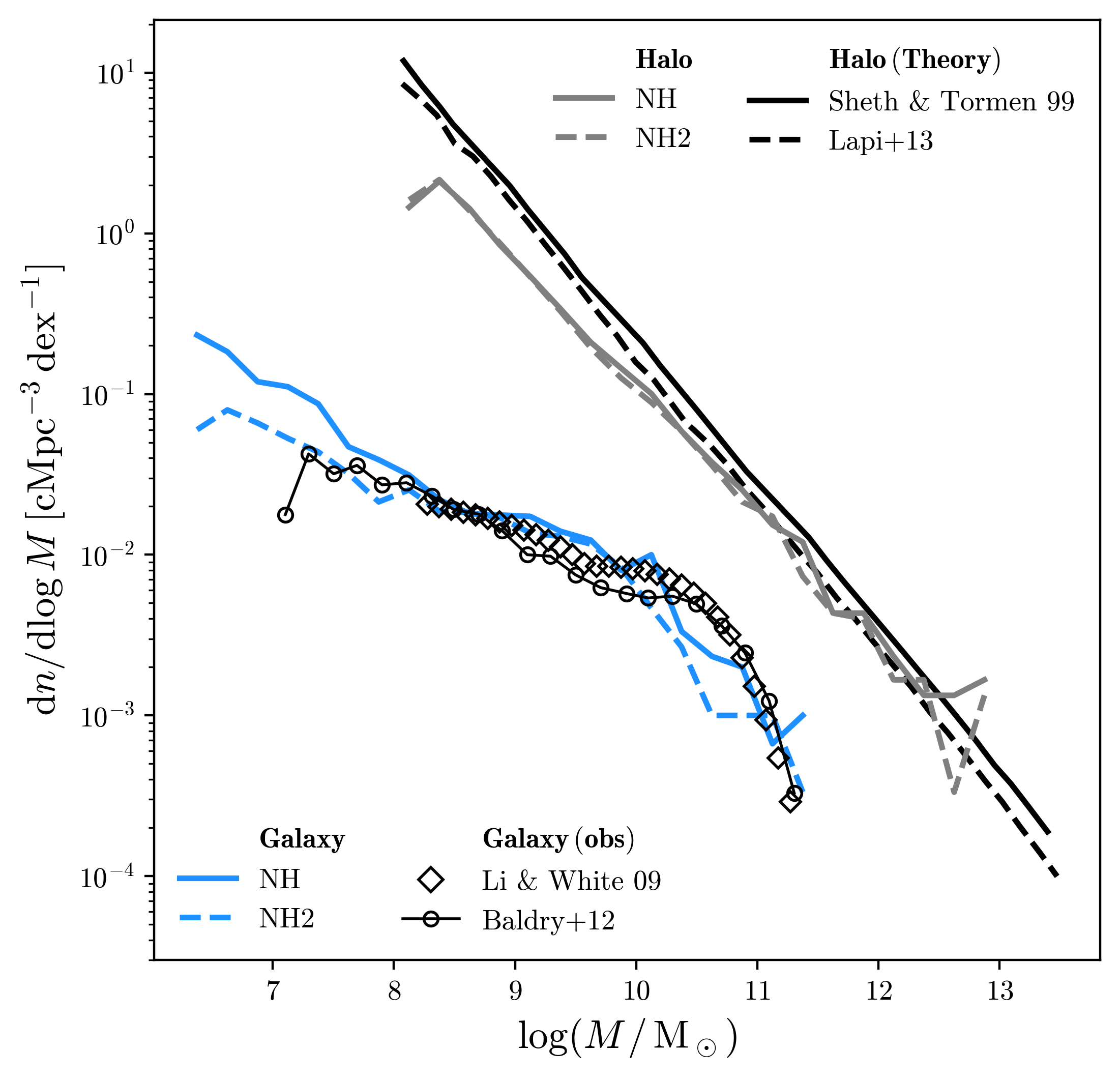}
    \caption{
        Mass functions of identified galaxies (blue) and halos (grey) in \NH\ and \NHII\ at $z\approx0.17$.
        The observed galaxy stellar mass functions are represented by black circles and diamonds \citep{Li2009, Baldry2012}.
        Theoretical halo mass functions are shown in black solid and dashed lines \citep{Sheth1999, Lapi2013}.
        }
    \label{fig01_HMF}
\end{figure}
%=================================================================%
%=================================================================%

%%%%%%%%%%%%%%%%%%%%%%%%%%%%%%%%%%%%%%%%%%%%%%%%%%%%%%%%%%%%%%%%%%%%%%%%%%%
%%%%% 	[Halo identification]
%%%%%%%%%%%%%%%%%%%%%%%%%%%%%%%%%%%%%%%%%%%%%%%%%%%%%%%%%%%%%%%%%%%%%%%%%%%
\subsection{Halo (galaxy) identification and sampling}\label{sec_method_sample}

% =================================================================
% > HaloMaker
The identification of galaxies and their associated halos is an important task in the context of this work.
Therefore, we provide here a detailed description of our identification method.
We separately identify DM halos and galaxies using the AdaptaHOP algorithm based on the particle density \citep{Aubert2004, Tweed2009}.
The minimum particle number is set to 100, corresponding to the mass of $\sim10^6\,\msol$ for galaxies and $\sim10^8\,\msol$ for halos.
The numbers of identified galaxies and halos are 4,070 (1,528) and 29,333 (29,176) for \NH\ (\NHII), respectively.
The difference between the two simulations in the number of galaxies originates from the stellar mass resolution, which significantly affects the lowest mass of galaxies.
We find that the number of DM halos is preserved due to the same DM mass resolution.
The large discrepancy between the numbers of galaxies and halos arises not only from the regulated star formation but also from the mechanism of the zoom-in technique.
By definition, zoom-in simulations prohibit star formation outside the zoom-in region, leaving only low-resolution DM and gas cells in those outer regions.
We exclude contaminated halos from our analysis.
We also present the mass function of the halos and galaxies with the comparison of observational results \citep{Li2009, Baldry2012} in Figure~\ref{fig01_HMF}.
The galaxy stellar mass functions from our simulations align well with observations, though the highest mass bins show fluctuations due to a lack of samples.

%=================================================================%
%          [Figure 02]
%=================================================================%
\begin{figure*}[htb!]
    \centering
    \includegraphics[width=0.90\textwidth]{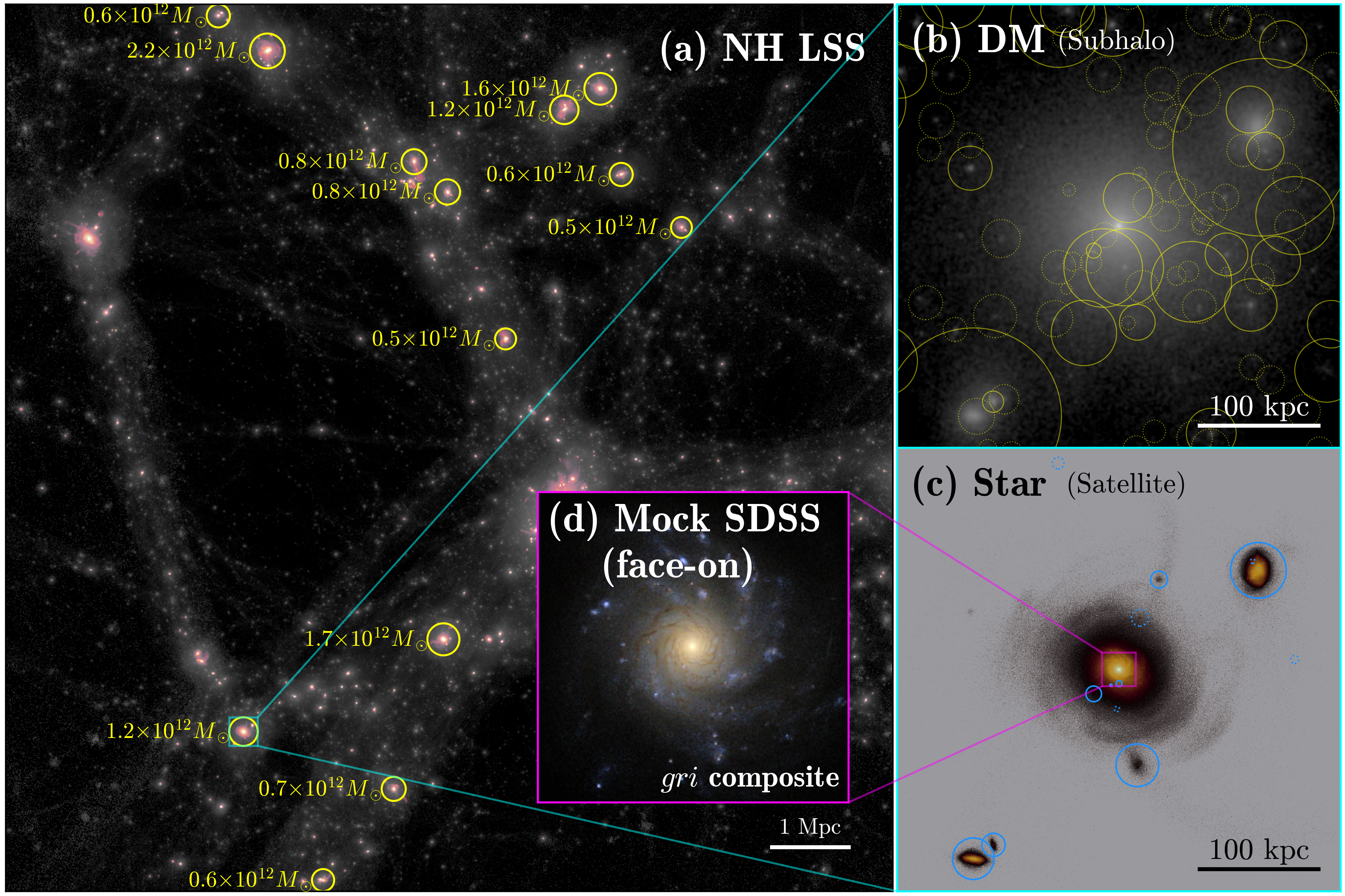}
    \caption{
        Overview of MWA systems and satellite statistics in our simulations.
        Panel (a) shows the large-scale distribution of DM (greyscale) and stars (reddish) in part of \NH.
        The identified MWA systems are marked with yellow circles, indicating the virial radius $R_{200,c}$ and mass $M_{200,c}$.
        Panel (b) zooms into a region within $R_{200,c}$ of one example system, illustrating the distribution of DM and subhalos.
        Yellow circles are identified subhalos in this system, with sizes corresponding to their virial radii.
        Solid circles represent more massive subhalos ($M_{\rm vir}>5\times10^8\,\msol$), while dotted circles represent less massive ones.
        Panel (c) shows the stellar distribution for the same system, with blue circles marking the identified satellite galaxies.
        Similarly, the line style distinguishes satellites with stellar masses above (solid) or below (dotted) $5\times10^6\,\msol$.
        Panel (d) displays its mock SDSS $gri$ composite image viewed face-on.
        }
    \label{fig02_example}
\end{figure*}
%=================================================================%
%=================================================================%

% =================================================================
% > MW and MWA analogs
We then selected Milky Way analogs.
% Bignone2019: EAGLE / VanNest2023: Romulus25 / Pillepich2024: TNG50
Even though many different criteria are exercised to define Milky Way analogs \citep{Bignone2019, VanNest2023, Pillepich2024}, we apply a simple and generous mass criterion since our simulations do not contain many massive halos.
We adopt the mass estimation from the observations: the central galaxy mass of $6.08\pm1.14\times10^{10}\,\msol$ \citep{Licquia2015}.
We double the mass range (e.g., $2\sigma$) to secure more samples.
To ensure that the host halo mass is around $\sim10^{12}\,\msol$ \citep{Callingham2019}, we exclude samples whose host halo mass lies outside the logarithmic range $[11.65, 12.35]$.
In total, 26 Milky Way analogs are selected in \NH\ and \NHII\ (13 in each simulation) at the last snapshot ($z=0.17$).
We define Milky Way analog systems (MWA systems) as the DM halos hosting the selected Milky Way analogs.

% =================================================================
% > Substructures
To investigate the subsystems (i.e., subhalos and satellite galaxies), we find all subhalos and satellite galaxies within $1.5R_{\rm 200,c}$ in the MWA systems, where $R_{\rm 200,c}$ is the radius within which the enclosed density is 200 times the critical density.
Of them, we want to exclude any misidentified substructures that are likely transient, such as star-forming clumps or fragmented structures.
To do this, we track the progenitors for the past 1 Gyr and calculate the score based on the number of shared member particles.
Using this merit score, we exclude all galaxies or halos that suddenly appear (fragmented) or disappear (transient).

% =================================================================
% > Example with Figure02
Figure~\ref{fig02_example} shows an example image of MWA systems in \NH.
We emphasize that these systems evolve in cosmological environments along large-scale filaments (panel (a)), and the sub-kpc scale structures, such as spiral arms or dwarf satellites, are resolved for individual galaxies (panel (d)).
% =================================================================
% > Example of the MWA system in NH
In panels (b) and (c), we show the zoom-in images of one example system with a box size of 2$R_{200,c}$ and mark identified subhalos and satellite galaxies on the DM and stellar density maps.
In line with recent studies \citep{Wadepuhl2011, Wetzel2016, Lee2024}, the satellite galaxies in panel (c) are far fewer than the subhalos in panel (b).
In panel (d), we also present the mock SDSS $gri$ composite image of the central MWA using the Monte Carlo radiative transfer code, SKIRT \citep{baes2015SKIRT, Camps2020}.

%%%%%%%%%%%%%%%%%%%%%%%%%%%%%%%%%%%%%%%%%%%%%%%%%%%%%%%%%%%%%%%%%%%%%%%%%%%
%%%%% 	[Starred & Starless classification]
%%%%%%%%%%%%%%%%%%%%%%%%%%%%%%%%%%%%%%%%%%%%%%%%%%%%%%%%%%%%%%%%%%%%%%%%%%%
\subsection{Classification of subhalos}\label{sec_method_classify}

% =================================================================
% > Two types of subhalos classification 
We now classify subhalos into two populations: ``starless'' and ``starred.''
Starless subhalos are subhalos without a galaxy, and starred subhalos are those with one.
However, since we identified the halos and the galaxies separately, we must match them.
Firstly, we identify the galaxy candidates hosted by each subhalo.
Ideally, we expect that the galaxy should be located at the center of the host halo, which is generally true in most cases mainly for massive and isolated galaxies.
However, especially for low-mass galaxies in dense environments, which are the main targets of this work, the galaxy could be offset from the center of the host halo or even outside the virial radius.
For instance, when a small subhalo with a galaxy orbits in a larger halo potential, the distribution of DM can get diffused, and its virialized region can shrink, while the galaxy maintains its distribution.
This could lead to an offset between the centers of the subhalo and the galaxy, which makes it hard to detect their connection simply based on catalog data.
Because of these offsets, we use the following criteria to ensure a robust association between halos and galaxies:
\begin{enumerate}
\item The center of the halo is in the galaxy: $(d < R_{\rm gal}) \land (R_{\rm gal} < R_{\rm hal})$,
\item The galaxy is in the halo: $d < (R_{\rm hal, vir}-R_{\rm gal})$,
\item The galaxy and the halo are overlapping $(d < 0.5\times(R_{\rm gal}+R_{\rm hal, vir})) \land (R_{\rm gal} < R_{\rm hal, vir})$,
\end{enumerate}
where $d$ is the distance between the center of the halo and the galaxy, $R_{\rm gal}$ and $R_{\rm hal}$ are the radii of the galaxy and the halo, respectively, and $R_{\rm hal, vir}$ is the virial radius of the halo.
We note that our halo finder, HaloMaker, provides a quantity `\texttt{r}' which represents the maximum distance between the center and the farthest member particle.
Another quantity, the virial radius `\texttt{rvir}', is determined by satisfying the virial theorem through direct computation.
$R_{\rm gal}$ and $R_{\rm hal}$ indicate an \texttt{r} to represent spatial occupation based on the maximum distance of member particles.
$R_{\rm hal, vir}$ is an \texttt{rvir}  which is a boundary of virialization.
Stellar, cold gas, and total gas masses in the main text are measured within \texttt{rvir}.
If a halo does not satisfy the first criterion (i.e., no matched galaxy), we sequentially check the second and third criteria.
We then classify subhalos as starless or starred based on the presence of a matched galaxy.
If there are multiple galaxy candidates, we select the galaxy closest to the center of the halo.

\begin{figure}[htb!]
    \centering
    \includegraphics[width=0.40\textwidth]{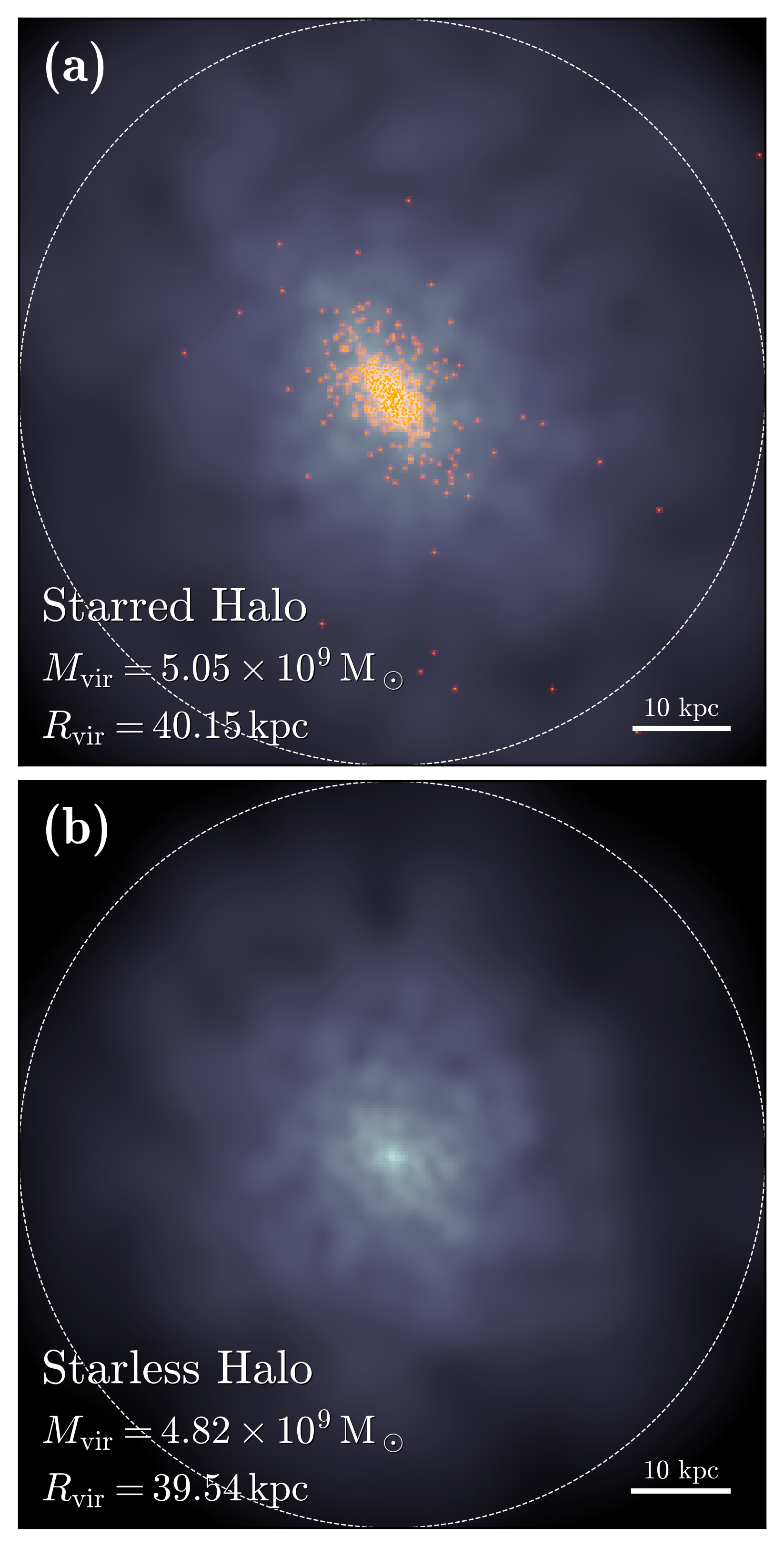}
    \caption{
        Example starless and starred subhalos in our systems at $z=0.17$, the final snapshot.
        Panels (a) and (b) show the images of sample starred and starless subhalos, respectively.
        The DM particles are shown as a smoothed density map in the background, and the stellar particles are overlaid in orange.
        White circles indicate $R_{\rm vir}$ computed from the halo finder.
        }
    \label{fig03_starless_example}
\end{figure}

% =================================================================
% > Further consideration
After this procedure, we track their merger trees to validate their classified labels.
We leave the detailed process of building the merger tree and further classification in Appendix~\ref{sec_method_tree} and \ref{sec_validity}, respectively.
Finally, we have 2,032 starless subhalos and 416 starred subhalos.
Representative images of the two types of subhalos at $z=0.17$ are shown in Figure~\ref{fig03_starless_example}.
Despite their similar virial masses, only the starred subhalo hosts stars (orange).

%=================================================================%
%          [Figure 03]
%=================================================================%
%=================================================================%
%=================================================================%

%%%%%%%%%%%%%%%%%%%%%%%%%%%%%%%%%%%%%%%%%%%%%%%%%%%%%%%%%%%%%%%%%%
%  ______    _______  _______  __   __  ___      _______  _______ 
% |    _ |  |       ||       ||  | |  ||   |    |       ||       |
% |   | ||  |    ___||  _____||  | |  ||   |    |_     _||  _____|
% |   |_||_ |   |___ | |_____ |  |_|  ||   |      |   |  | |_____ 
% |    __  ||    ___||_____  ||       ||   |___   |   |  |_____  |
% |   |  | ||   |___  _____| ||       ||       |  |   |   _____| |
% |___|  |_||_______||_______||_______||_______|  |___|  |_______|
%%%%%%%%%%%%%%%%%%%%%%%%%%%%%%%%%%%%%%%%%%%%%%%%%%%%%%%%%%%%%%%%%%
\section{Results}
\label{sec_results}

% =================================================================
% > Opening results section
In this section, we present the main results of our analysis, demonstrating that the so-called missing satellite problem does not arise in our simulations, as the majority of subhalos do not host a galaxy.
Building on this finding, we then further explore the underlying reasons why this population of starless subhaloes remains devoid of galaxies, examining the physical processes that may be responsible for their evolution and fate within the host halos.

%=================================================================%
%          [Figure 04]
%=================================================================%
\begin{figure}[htb!]
    \centering
    \includegraphics[width=0.45\textwidth]{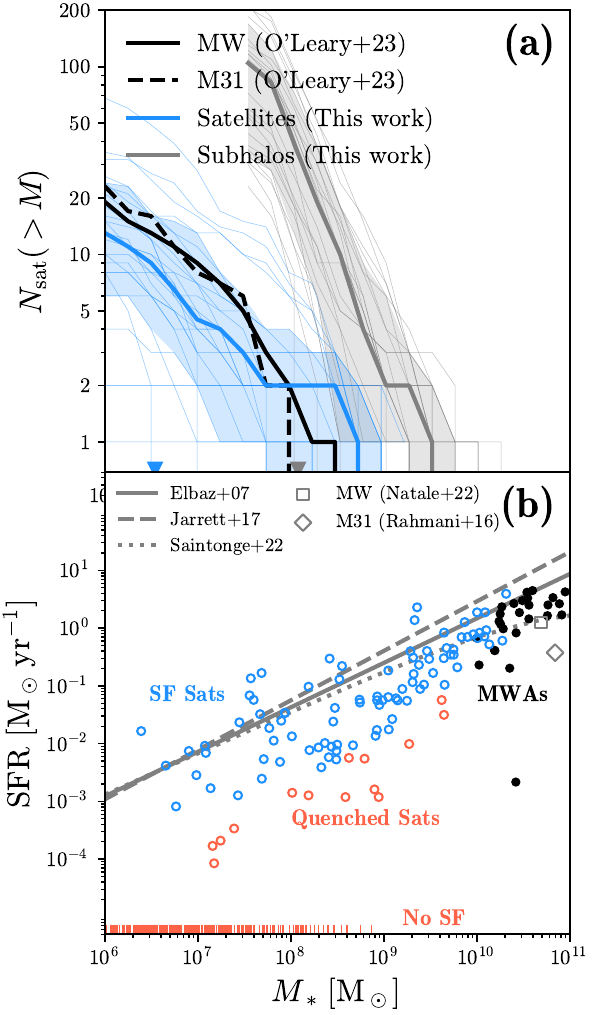}
    \caption{
        Number and star formation rates of satellite galaxy samples.
        Panel (a) shows the cumulative number of satellite galaxies (blue) and subhalos (grey) in our MWA systems in \NH\ and \NHII, plotted with respect to the stellar mass of satellites and the virial mass of subhalos, respectively.
        Arrow markers at $5\times10^6\,\msol$ (galaxy) and $5\times10^8\,\msol$ (halo) correspond to $\sim500$ particles.
        Thicker lines are the median, and shaded regions are the 16$^{\rm th}$ to 84$^{\rm th}$ percentiles.
        Black lines represent observed values for the Milky Way and M31 (stellar mass extracted from Table A1 of \citealt{O'Leary2023}).
        Panel (b) shows the star formation rate and the stellar mass of the satellite galaxies.
        The blue circles exhibit the star-forming satellites, the red circles are the quenched satellites, the red triangles are those with zero SFR, and the black dots are the Milky Way analogs in our simulations.
        The grey lines are observational star-forming main sequence \citep{Elbaz2007, Jarrett2017, Saintonge2022}, and the grey squares are the observed values of the Milky Way and M31 \citep{Natale2022, Rahmani2016}.
        }
    \label{fig04_number}
\end{figure}
%=================================================================%
%=================================================================%

%%%%%%%%%%%%%%%%%%%%%%%%%%%%%%%%%%%%%%%%%%%%%%%%%%%%%%%%%%%%%%%%%%%%%%%%%%%
%%%%% 	[No Missing Satellite Problem]
%%%%%%%%%%%%%%%%%%%%%%%%%%%%%%%%%%%%%%%%%%%%%%%%%%%%%%%%%%%%%%%%%%%%%%%%%%%
\subsection{Satellite abundance and star formation}
\label{sec_missing_sat}

% =================================================================
% > Number of satellite galaxies
We present the cumulative number of satellite galaxies and subhalos as a function of mass in panel (a) of Figure~\ref{fig04_number}.
Statistically, one can see clearly in panel (a) that the sequences of satellite galaxies (blue) and subhalos (grey) are well separated.
Especially for our satellites, the cumulative numbers (blue line) match well with the observed numbers of Milky Way and M31 satellites (black lines) of \citet{O'Leary2023}.
We note that recent surveys, such as DES, Pan-STARRS, and SAGA, have extended detection to fainter satellites or to other galaxy systems \citep{Nadler2020, Mao2021, Smercina2018, Nashimoto2022}.
We decided to use only the Local Group sample.
It is, first of all, because different galaxy data show different star formation properties \citep{Smercina2018, Mao2021}. 
Meanwhile, the Local Group sample is sufficient to tell us that there are far more starless subhalos than starred ones in the real Universe, which is the main motivation for our study. 
On a different account, we did not use fainter satellite observations than those used in this study, because our simulations do not reach so faint anyway.
This reproduction of satellite abundance aligns with results reported previously from other models \citep{Wetzel2016, Sawala2016, Engler2021, Lee2024, Jung2024}, further supporting the alleviation of the classical missing satellite problem for this stellar mass range.

% =================================================================
% > Star formation properties
The star formation properties of our satellite samples are also aligned with the observational data.
In panel (b), we measure the star formation rate of our samples and observational results and classify satellite galaxies as either star-forming or quenched using the specific star formation rate cut \citep{Tacchella2019}.
We find that the star-forming satellites (blue circles) are well located on the star-forming main sequence \citep[grey lines;][]{Elbaz2007, Jarrett2017, Saintonge2022}.
The trend that a significant fraction of the satellites are quenched (red circles) is also well matched with the observed quenched fraction in Local group \citep{Wetzel2015}, although it is much higher than the result of the SAGA survey \citep{Mao2021}.

Since our samples seem to agree largely with the general understanding of not only abundance but also star formation properties, we now want to focus on the individual subhalos to understand their stellar contents.

%=================================================================%
%          [Figure 05]
%=================================================================%
\begin{figure*}[htb!]
    \centering
    \includegraphics[width=0.95\textwidth]{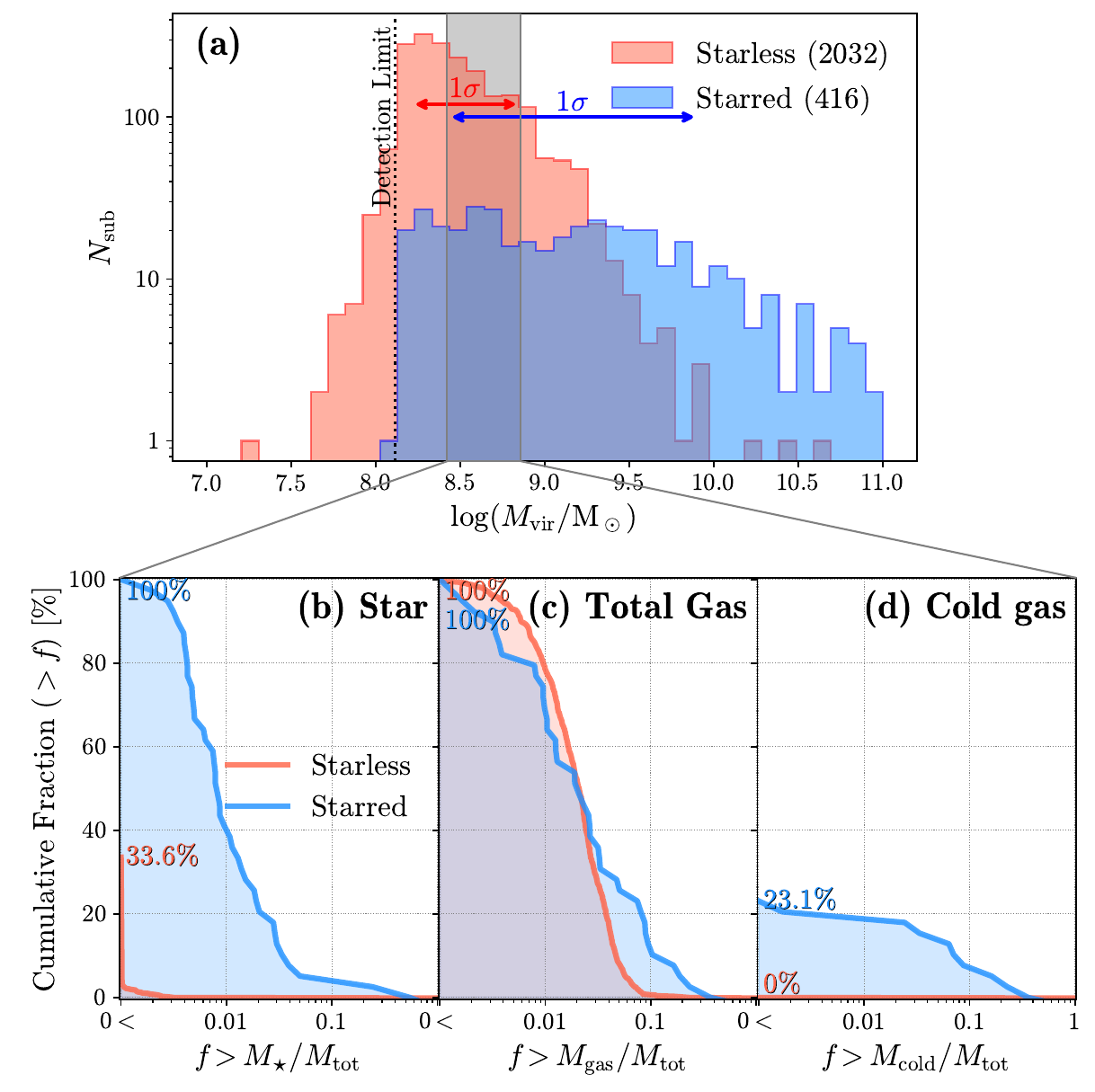}
    \caption{
        Structural and compositional differences between starless and starred subhalos in our systems at $z=0.17$, the final snapshot.
        Panel (a) shows their mass functions.
        Arrows in the distribution indicate the $1\,\sigma$ ranges.
        The overlapping range (grey band) of $10^{8.4} \lesssim M_{\rm vir}/\msol \lesssim 10^{8.9}$ is used to select subhalos for analysis in panels (b), (c), and (d).
        The black dotted vertical line indicates the minimum threshold of halo identification ($1.3\times10^8\,\msol$).
        Panels (b), (c), and (d) display the cumulative fractional distribution of stars, total gas, and cold gas, respectively.
        The fraction of subhalos containing each component is also shown.
        Note that the starless subhalos have comparable amounts of gas (panel (c)) but lack stars (panel (b)), probably because their gas was not cold and dense enough (panel (d)).
        }
    \label{fig05_z0diff}
\end{figure*}
%=================================================================%
%=================================================================%

%%%%%%%%%%%%%%%%%%%%%%%%%%%%%%%%%%%%%%%%%%%%%%%%%%%%%%%%%%%%%%%%%%%%%%%%%%%
%%%%% 	[Starless and Starred]
%%%%%%%%%%%%%%%%%%%%%%%%%%%%%%%%%%%%%%%%%%%%%%%%%%%%%%%%%%%%%%%%%%%%%%%%%%%
\subsection{Distinct subhalo populations: starless and starred}
\label{sec:sub_populations}

% =================================================================
% > The existence of the starless and the starred
The effect of baryons seems to play a crucial role in resolving the missing satellite problem.
Their impact extends beyond just reducing the stellar mass of satellite galaxies, but it even influences their existence.
From the large discrepancy in mass functions of subhalos and satellite galaxies, we see that only a few subhalos host galaxies.
We name the subhalos without significant stellar components as ``starless'', and those with prominent satellite galaxies as ``starred'', as mentioned in Section~\ref{sec_method_classify}.\footnote[1]{A main criterion to separate them is sustainability to host stars.
Even if there are some stellar particles within the subhalo in question, we classify it as a starred subhalo only when it continues to contain the stars for enough time, excluding the temporary invasion from neighboring galaxies.
Detailed procedure is explained in the Section~\ref{sec_validity}.}

% =================================================================
% > Different Mass distribution and common mass range
In Figure~\ref{fig05_z0diff}, the statistical comparison of the two subhalo types reveals a clear difference in their overall mass distribution (panel (a)).
In general, starred subhalos have larger masses than starless subhalos.
More massive halos tend to attract more gas, leading to increased star formation and the development of starred subhalos.
For the lowest mass subhalos, some of them barely exceed the minimum identification threshold, making them unreliable.
Nevertheless, there is a range of common mass between the two populations.
We select the subhalos of $M_{\rm vir}=10^{8.4\,\text{--}\,8.9}\,\msol$ where the $1\,\sigma$ ranges of the two mass functions overlap.
We now examine the subhalos in this common mass range to find what makes a subhalo starred or starless, setting aside the halo mass effect.

% =================================================================
% > Different components in the starless and the starred
We measure the fractional mass distribution of subhalo components, such as stars, total gas, and cold gas (below $10^4\,\rm{K}$, regardless of hydrogen species).
Panel (b) shows the dramatic difference in stellar content between the two subhalo types.
A small fraction of starless halos have stars (red peak near $f>0$).
In most cases, these stars are interlopers that temporarily pollute these subhalos from larger neighbor galaxies.
Despite this large difference, the two populations contain a similar amount of gas (panel (c)), which is very amusing.
The two populations are well separated in the cold gas content (panel (d)).
Starless subhalos completely lack cold gas, indicating that star formation is totally absent.
Then, the key question is whether starless subhalos have ever hosted galaxies or if they have lost them during their evolution.

% =================================================================
% > Evolution: why do they lack cold gas?
We examine two of the most widely-discussed mechanisms that may explain the lack of cold gas during the evolution of starless subhalos: supernova feedback and environmental effects associated with the orbital motion of subhalos.
These two are representative processes for low-mass galaxy evolution and quenching, namely ``nature or nurture'' \citep{Peng2010, Balogh2016, Darvish2016, kawinwanichakij2017, Rhee2020, Contini2020a}.
The {\em final} mass of a subhalo does not necessarily reflect its entire growth history.
Therefore, to minimize the influence of different assembly histories, we sample subhalos with similar evolution using {\em peak} masses.\footnote[2]{We use the same method previously described for final mass binning (Figure~\ref{fig05_z0diff}(a)) to determine the common mass range between starred and starless subhalos based on their peak masses.
We apply a mass range between the 8$^{\rm th}$ to 92$^{\rm nd}$ percentiles to ensure a sufficient number of samples.}
Using these subsamples, we first investigate the significance of the two candidate processes mentioned above for removing cold gas.

%=================================================================%
%          [Figure 06]
%=================================================================%
\begin{figure*}[htb!]
    \centering
    \includegraphics[width=0.95\textwidth]{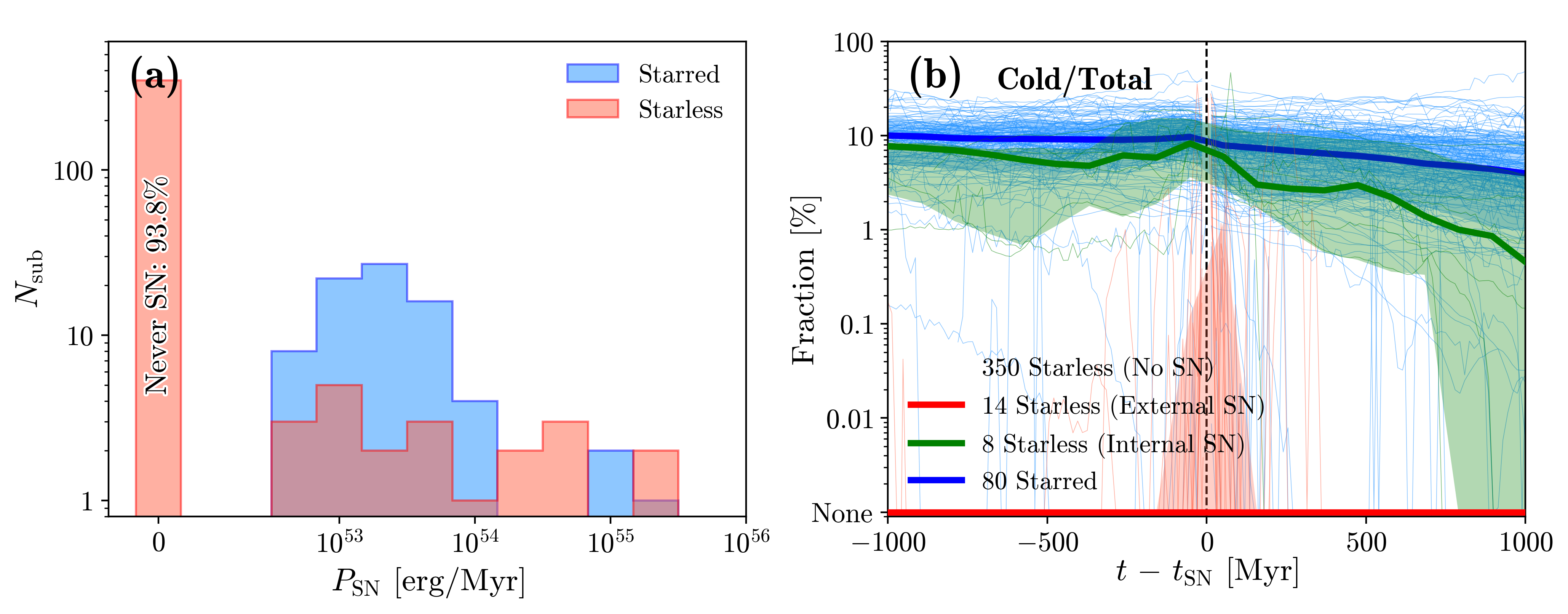}
    \caption{
        The first main possible scenario responsible for the removal of the cold gas in certain halos: supernova feedback.
        Panel (a) shows the supernova energy rate distribution from the starred (blue) and starless (red) subhalos.
        We also present the percentage of starless subhalos that do not have any supernova events (93.8\%).
        Panel (b) exhibits the time evolution of the cold gas mass fraction, with the x-axis representing the time offset from the supernova explosion.
        The black vertical line indicates the epoch of supernova explosion.
        Thin solid lines are the evolution of individual subhalos, while thick lines and shaded regions represent the median values and the 16$^{\rm th}$ to 84$^{\rm th}$ percentiles, respectively.
        Colors represent different populations of supernova-hosting subhalos: blue for starred subhalos, green for starless subhalos with internal supernovae, and red for starless subhalos with external supernovae from neighboring halos.
        }
    \label{fig06_SN}
\end{figure*}
%=================================================================%
%=================================================================%

%%%%%%%%%%%%%%%%%%%%%%%%%%%%%%%%%%%%%%%%%%%%%%%%%%%%%%%%%%%%%%%%%%%%%%%%%%%
%%%%% 	[Supernova Feedback]
%%%%%%%%%%%%%%%%%%%%%%%%%%%%%%%%%%%%%%%%%%%%%%%%%%%%%%%%%%%%%%%%%%%%%%%%%%%
\subsubsection{Supernova feedback}

% =================================================================
% > Evolution: Supernova feedback
First, supernova feedback is expected to be effective for removing cold gas through energy and momentum injection.
The surrounding gas can be heated by the release of supernova energy and, if loosely bound, expelled from the subhalo, resulting in the suppression of star formation \citep{Larson1974, Dekel1986, Dubois2008, Kimm2018}.
To investigate whether supernova feedback had a more significant impact on starless subhalos, we measure $P_{\rm SN}$, the expected value of supernova energy for any timestep when supernovae occur in Figure~\ref{fig06_SN}(a).
Detailed calculation is explained in Appendix~\ref{sec_snpower}, but for instance, a high value of $P_{\rm SN}$ indicates that a burst of supernovae occurred over a short period of time.

% =================================================================
% > SN in Starred subhalos -> All but not significant
As visible in panel (a), all the starred subhalos have large values of $P_{\rm SN}$.
In order to check its impact on cold gas contents, we show the cold gas fraction before and after the supernova events (vertical line) in panel (b).
The cold gas mass indeed slightly decreases after the supernova explosion in the starred subhalos (blue), as expected.
Nevertheless, these subhalos succeed in retaining a significant amount of cold gas afterward.
Supernova feedback probably does not affect existing stars of starred subhalos, but it may affect their growth by reducing the future star formation rates.

% =================================================================
% > Majority of starless -> No SN ever
Another striking feature in panel (a) is that most (93.8\%) of the starless subhalos do not experience any supernova events.
In fact, they do not have any in-situ star formation.
Therefore, supernova feedback is irrelevant as cold gas removal process in most starless subhalos.

% =================================================================
% > External SN contamination -> Similar to NoSN case
Some starless subhalos have comparable values of $P_{\rm SN}$ to starred subhalos in panel (a).
Before examining them, we would like to remind the reader that some subhalos can be influenced by supernovae from neighboring halos.
To account for this, we classify starless subhalos with supernova events into internal and external cases based on the true ownership of young stars.
When inspecting the star formation histories, we find that starless subhalos with external supernovae alone (N=14) have indeed not formed their own stars, as most starless subhalos.
Panel (b) shows that they (red line at $y=0$) had no cold gas, regardless of supernova events.
Thus, the external supernova case is also irrelevant as a test for the supernova feedback effect.

% =================================================================
% > Rare internal SN -> Could be, but rare
A small number (N=8) of starless subhalos have internal supernova events.
As shown in panel (b), they lose cold gas after their supernova events, and their cold gas decays faster than their starred counterparts.
These eight out of 372 starless subhalos (2\%) could be said to have become starless due to supernova feedback but are too few to represent the starless subhalos.

In summary, supernova feedback is not the main driver of cold gas removal in starless subhalos.

%%%%%%%%%%%%%%%%%%%%%%%%%%%%%%%%%%%%%%%%%%%%%%%%%%%%%%%%%%%%%%%%%%%%%%%%%%%
%%%%% 	[Environmental Effects]
%%%%%%%%%%%%%%%%%%%%%%%%%%%%%%%%%%%%%%%%%%%%%%%%%%%%%%%%%%%%%%%%%%%%%%%%%%%
\subsubsection{Environmental effects associated with orbital motion}

%=================================================================%
%          [Figure 07]
%=================================================================%
\begin{figure*}[htb!]
    \centering
    \includegraphics[width=0.95\textwidth]{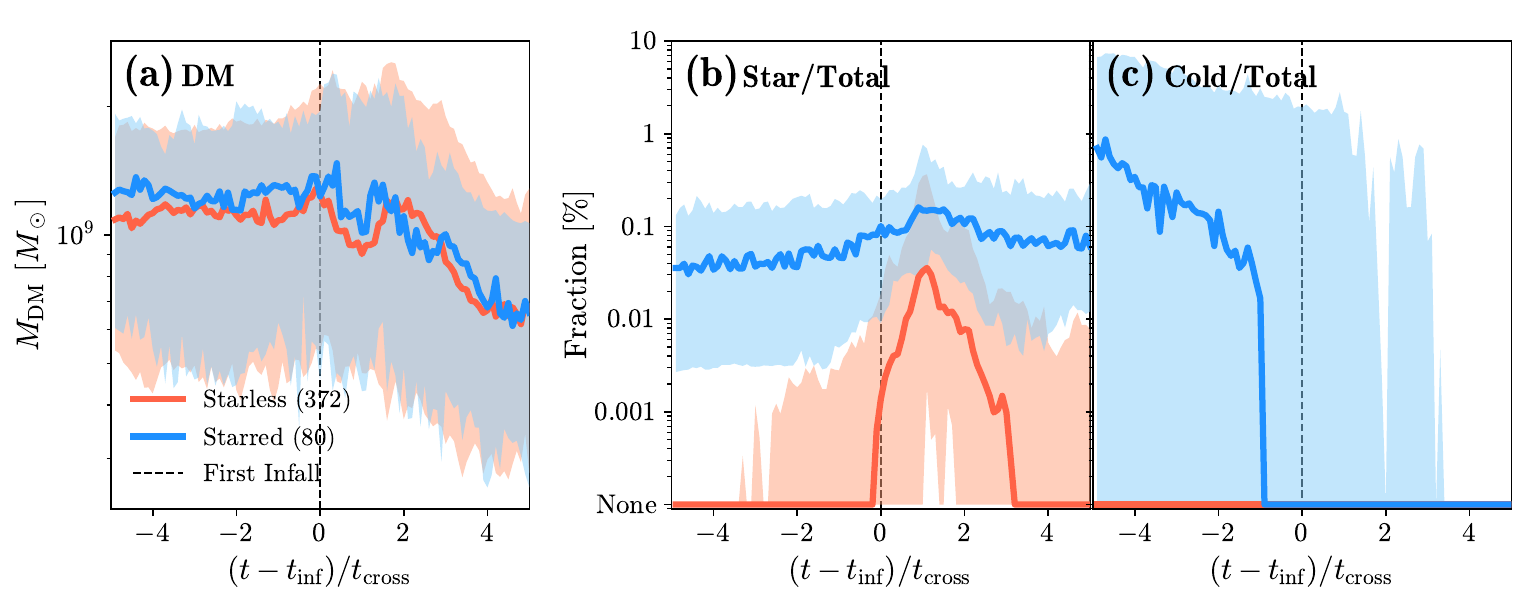}
    \caption{
        The second main possible scenario responsible for the removal of the cold gas in certain halos: the environmental effects associated with the orbital motion.
        Panel (a) illustrates the evolution of DM mass around the first infall into the proto-MWA systems.
        The x-axis is the time offset from the first infall, normalized by the crossing time.
        The black vertical line marks the first infall epoch.
        Panel (b) shows the stellar mass fraction, and panel (c) displays the cold gas fraction.
        }
    \label{fig07_env}
\end{figure*}
%=================================================================%
%=================================================================%

% =================================================================
% > Evolution: Environmental effects
We now investigate the orbital motion-associated environmental effects on subhalos.
Specifically, Local Group environments could remove cold gas of satellite subhalos through ram pressure stripping or even directly strip stars through tidal stripping \citep{Gunn1972, Richstone1976, Bahe2015, Contini2024, Rhee2024}, potentially driving their transition to starless subhalos.
To test this, we examine the evolution of their components using the time offset from the first infall normalized by the crossing time.
In this normalized unit, key orbital stages, such as first infall (at 0), pericenter passage (at 1), and apocenter passage (at 2), are conspicuous.
% (DM Change)
Figure~\ref{fig07_env}(a) shows the DM mass evolution, which is similar across the two populations, as expected from similar assembly histories of the sampled subhalos.
Before infalling into the MWA system, both populations exhibit a similar flat trend.
Following the first infall, both experience a decrease in DM due to tidal stripping.
We present the stellar and cold gas fractions relative to the total mass in panels (b) and (c), respectively.

% =================================================================
% > Different between two populations
% (Starless: Inappropriate for star)
It is inappropriate to discuss the environmental effects on starless subhalos because they hardly had any {\em cold} gas or stars in the first place.
The peak at the first pericenter passage ($x\approx1$) in panel (b) shows a transient feature of having interloper stars from neighboring halos, primarily from their central Milky Way analogs.
% (Starred: Not enough for star)
In the case of starred subhalos, their star fractions are not significantly affected by the orbital motion.
Aside from the transient bump near the first pericenter passage, the star fraction appears similar or even elevated.
This can be understood by the fact that DM stripping precedes stellar stripping due to differences in spatial distribution and concentration \citep{Contini2014, Smith2016, contini2023, Jang2024}.
So, the environmental effects are negligible on stars.
% (Cold gas)
On the other hand, the cold gas of starred subhalos is dramatically reduced through the infall due to the ram pressure stripping (panel (c)).
This explains why starred subhalos are largely deficient in cold gas in the final stage (Figure~\ref{fig05_z0diff}(d)).
In conclusion, for the case of starred subhalos, environmental effects are effective on cold gas, removing it quickly and quenching star formation, but do not remove pre-existing stars.
As a result, they cannot transform starred halos into starless ones.

% =================================================================
% > Concluding summary
We conclude that environmental effects associated with subhalo's orbital motions are not responsible for the absence of stellar components in starless subhalos, either.
Then, what caused starless halos to be starless?

%=================================================================%
%          [Figure 08]
%=================================================================%
\begin{figure}[htb!]
    \centering
    \includegraphics[width=0.45\textwidth]{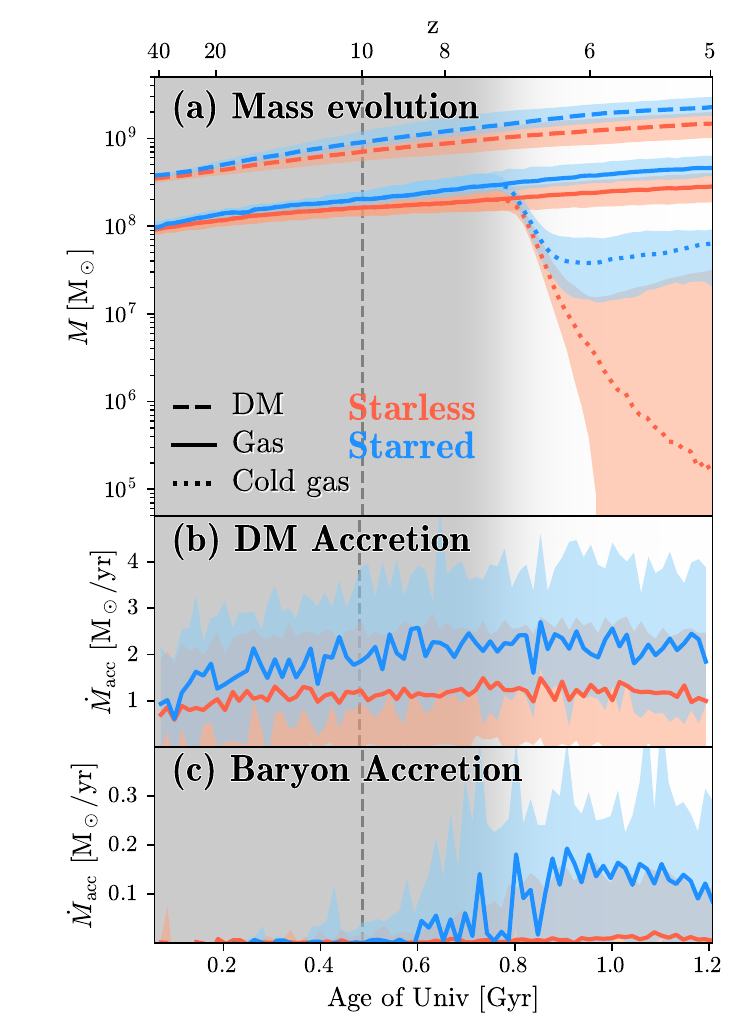}
    \caption{
        Comparison of the initial evolution between starred (blue) and starless (red) subhalos, measured within a 100 comoving kpc box centered on their birthplace.
        Panel (a) shows the median mass evolution of DM (dashed), gas (solid), and cold gas (dotted).
        Shaded regions are the 16$^{\rm th}$ to 84$^{\rm th}$ percentiles for each ingredient.
        The grey vertical line indicates the onset of reionization in our models.
        To visualize the end of reionization, we color the background of panels in greyscale (e.g., a white background indicates a fully ionized state).
        We extract the hydrogen ionization fraction for a representative gas cell ($10^4$ K and \hcc{0.001}) to illustrate the progression of reionization.
        Panel (b) shows the median evolution of DM accretion rates in the {\em box}, which indicates the amount of matter gathering on a large scale.
        Panel (c) is similar to panel (b) but shows baryon accretion rates in {\em virialized regions}.
        }
    \label{fig08_accretion}
\end{figure}
%=================================================================%
%=================================================================%

%%%%%%%%%%%%%%%%%%%%%%%%%%%%%%%%%%%%%%%%%%%%%%%%%%%%%%%%%%%%%%%%%%%%%%%%%%%
%%%%% 	[Born to be starless]
%%%%%%%%%%%%%%%%%%%%%%%%%%%%%%%%%%%%%%%%%%%%%%%%%%%%%%%%%%%%%%%%%%%%%%%%%%%
\subsubsection{Born to be starless}

% =================================================================
% > No Star Formation <- No dense & cold enough gas
Star formation prescription in our simulations follows the gravo-turbulent scheme \citep{Federrath2012} that considers local gravity and thermal/turbulent pressure \citep{Dubois2021}.
These criteria are applied to dense gas cells with a density threshold of \hcc{10} (\hcc{5} for \NHII).
However, in most starless subhalos, these dense cells rarely exist for the following reasons.

% =================================================================
% > No dense & cold enough gas <- UV heating <- Cannot self-shield
UV background is the dominant mechanism preventing cooling and subsequent star formation in starless subhalos.
Since most starless subhalos are very low in (virial) mass, they are sensitive to UV heating \citep{Navarro1997, Gnedin2000, Pereira-Wilson2023}.
The uniform UV background in our simulations after $z=10$ heats and evaporates gas, therefore suppressing star formation (see the visualized temperature map in Figure~\ref{figA2_tempmap} in Appendix~\ref{sec_cooling}).
To overcome this heating, halos must in the first place have a sufficiently deep potential and host gas with a density of at least \hcc{0.01}, the self-shielding threshold \citep{Rosdahl2012, Dubois2021}.
Subhalos that sustain dense gas can enrich their gas metallicity through star formation, enabling additional metal-cooling.
While most starred subhalos contain dense and cold gas before reionization, starless subhalos fail to reach this density threshold.
Thus, their gas is rapidly heated from early times.

% =================================================================
% > Different birthplace -> Different accretion rates
The failure of starless subhalos to have dense gas may be related to the condition of birthplace before reionization.
To verify this, we investigate the differences in the long-range environment.
In Figure~\ref{fig08_accretion}, we measure various initial evolutions within a 100 comoving kpc box centered on the birthplace of each subhalo.
The masses inside the box are nearly identical at birth (panel (a)), so both populations appear to form in similar environments with comparable matter content.
However, the mass growth rates of starless subhalos are slightly lower than those of starred subhalos, leading to a divergence in total mass over time.
The birth sites of starred subhalos exhibit higher matter accretion rates (panels (b) and (c)), indicating that these regions gather matter more efficiently.

%=================================================================%
%          [Figure 09]
%=================================================================%
\begin{figure}[htb!]
    \centering
    \includegraphics[width=0.45\textwidth]{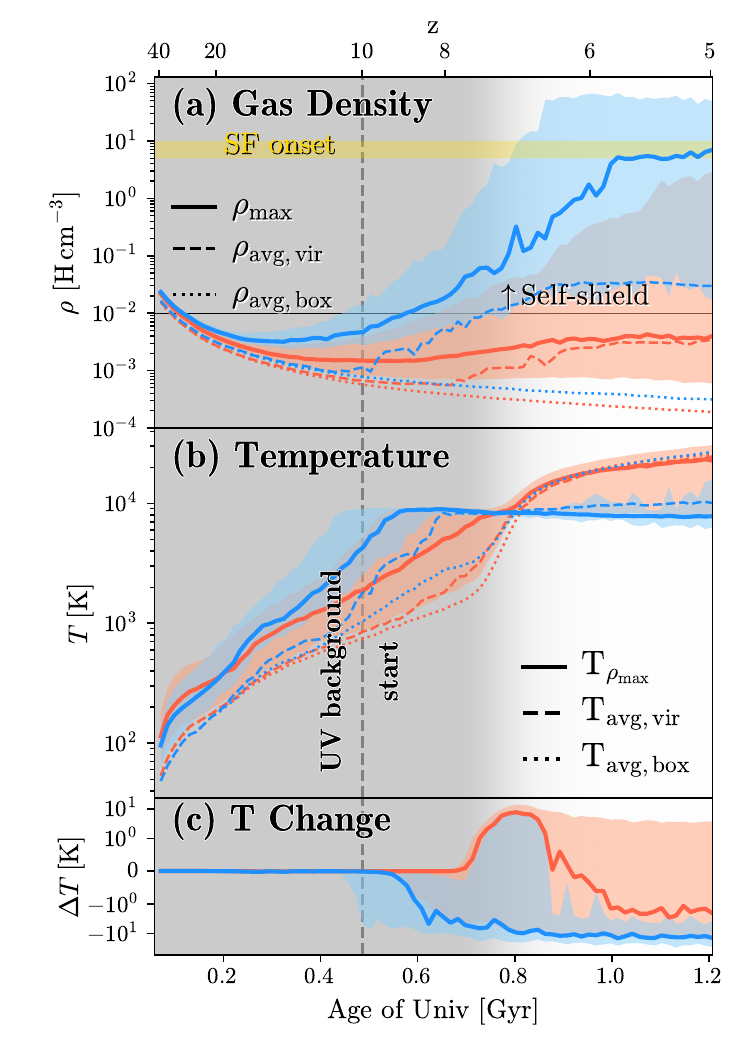}
    \caption{
        Similar evolutionary information to Figure~\ref{fig08_accretion}, but focused on gas.
        Panel (a) exhibits the evolution of the gas density for each population.
        Solid lines represent the maximum density, while dashed (dotted) lines are the average density within the virialized region (box).
        We also indicate the density thresholds for star formation (\hcc{10} for \NH\ and \hcc{5} for \NHII; yellow shades) and self-shielding (\hcc{0.01}; black horizontal line).
        Panel (b) is similar to panel (a) but depicts the evolution of temperature.
        Solid lines are the temperature of the densest gas cells, while dashed (dotted) lines are the average temperature in the virial region (box).
        Panel (c) shows the net temperature change of the top 1\% densest gas cells, derived only from the {\em radiative cooling and heating} rates.
        }
    \label{fig09_gasevol}
\end{figure}
%=================================================================%
%=================================================================%

% =================================================================
% > Different accretion rates -> Different gas density growth
Higher accretion rates in starred subhalos also drive rapid gas density growth (panel (a) of Figure~\ref{fig09_gasevol}).
Similar to bulk gas growth shown in Figure~\ref{fig08_accretion}(a), the average gas densities (dotted lines) of the two populations in Figure~\ref{fig09_gasevol}(a) start at similar levels and slowly diverge over time due to different growth rates.
The maximum gas densities (solid lines) show more dramatic divergence early on.
After the initial density decreases, both the maximum density and the average density in the {\em virialized} region (dashed lines) increase in starred subhalos.
Thanks to more efficient accretion, gas in starred subhalos can exceed the self-shielding density (i.e., \hcc{0.01}) before the reionization becomes prevalent at $z\sim7$.
This dense gas can resist UV heating, and further cool and collapse, eventually touching the star-forming density (i.e., \hcc{5-10}).
By contrast, starless subhalos show a stagnant level of gas density far below the self-shielding threshold.
They do not have dense enough gas to self-shield even through the completion of reionization.

% =================================================================
% > Different gas density -> Different fate is fixed at reionization
Reionization is completed at $z\sim7$ in our simulations, which is aligned with the sudden decrease of the cold gas mass in Figure~\ref{fig08_accretion}(a).
During the reionization transition period (i.e., $7<z<10$), the directional difference between the two populations becomes noticeable.
The temperature (Figure~\ref{fig09_gasevol}(b)) of the densest gas in starred subhalos peaks at $\rm \sim10,000 K$ at this period and then cools down marginally but steadily.
In contrast, starless subhalos show a gradual increase in temperature, and even for the densest cells, the trend of temperature evolution mirrors the average temperature in the box.
% > Detailed radiative effect
In Figure~\ref{fig09_gasevol}(c), we can see the impact of reionization more explicitly by measuring the gas temperature change caused only by radiative cooling and heating, excluding adiabatic effects.
The key difference begins near the onset of reionization ($z=10$).
Before the completion of reionization, gas in starred subhalos becomes dense enough to self-shield, suppressing radiative heating and enabling continuous cooling.
In contrast, gas in starless subhalos remains too diffuse (panel (a)) to self-shield, leading to inefficient radiative cooling.
After the UV background becomes fully prevalent, starless subhalos undergo a dominantly heating phase, while starred subhalos exhibit a strong net cooling phase (panels (b) and (c)).

% =================================================================
% > Concluding remarks
The fate of subhalos seems to be determined at birth.
The difference in birth sites may reflect variations in the subhalo's connection to the cosmic web (e.g., \citealt{Musso2018}).
While further analysis is required to clarify this point, the initial difference appears to be amplified and sealed during reionization.
Starless subhalos live their starless fate thereafter.

Building upon the analysis presented in this section, the next will be dedicated to a comprehensive discussion of our results, placing them in the broader context of previous
key studies on the same topic.
We will highlight the main similarities and differences between our findings and those reported in the literature, addressing potential reasons for any discrepancies.

%%%%%%%%%%%%%%%%%%%%%%%%%%%%%%%%%%%%%%%%%%%%%%%%%%%%%%%%%%%%%%
%  ______   ___   _______  _______  __   __  _______  _______ 
% |      | |   | |       ||       ||  | |  ||       ||       |
% |  _    ||   | |  _____||       ||  | |  ||  _____||  _____|
% | | |   ||   | | |_____ |       ||  |_|  || |_____ | |_____ 
% | |_|   ||   | |_____  ||      _||       ||_____  ||_____  |
% |       ||   |  _____| ||     |_ |       | _____| | _____| |
% |______| |___| |_______||_______||_______||_______||_______|
%%%%%%%%%%%%%%%%%%%%%%%%%%%%%%%%%%%%%%%%%%%%%%%%%%%%%%%%%%%%%%
\section{Discussion}
\label{sec_discussion}

%%%%% Previous works that highlight reionization
The results obtained in this analysis provide a detailed picture of the relationship between subhalo mass and their ability to host visible galaxies, contributing to our understanding of the properties of dwarf galaxies in the Local Group and the role of baryonic feedback in suppressing star formation in low-mass subhalos.
Our data suggest that the abundance and distribution of satellite galaxies are strongly influenced by feedback processes and tidal stripping, in agreement with the findings of \cite{Sawala2016}.
In that study, the authors demonstrated, through hydrodynamic simulations, that baryonic heating and reionization significantly reduce the final mass of subhalos, making them more vulnerable to tidal destruction.
This insight builds on earlier works, such as those by \cite{Efstathiou1992}, \cite{Thoul1996}, and \cite{Benson2002}, which proposed that reionization prevents gas from cooling and condensing in halos below a mass threshold determined by the virial temperature of the heated gas.
Our results confirm that including baryonic effects can explain the relatively low number of observed satellites in the Milky Way compared to the predictions of purely gravitational simulations.
However, as noted by \cite{Benitez2019} and \cite{Benitez2020}, this effect is compounded by gas cooling and dynamical instabilities, which further influence subhalo survival.

%%%%% Mass distribution and threshold (vs Simpson+18)
One of the most significant aspects emerging from our study concerns the mass distribution of subhalos that host luminous galaxies.
While previous work such as \citet{Simpson2018}—based on hydrodynamic simulations—suggested that the visibility of subhalos is shaped by reionization and star formation feedback, we find that the picture is more nuanced.
Thanks to the higher mass and spatial resolution of our simulations, we observe that subhalos with and without stars occupy a broad and overlapping mass range below the suggested critical mass threshold of $M_{\rm halo} \sim 10^9\,\msol$ (Figure~\ref{fig05_z0diff}).
This suggests that satellite survival is not governed by a sharp mass threshold.
Instead, the formation of luminous galaxies in low-mass subhalos is likely influenced by their early assembly history, not merely their halo mass at $z=0$.
% FIRE
This behavior has also been reported in zoom-in simulations by \citet{Fitts2017}, where late-forming and low-concentration halos exhibit totally absent stars despite reaching masses comparable to luminous counterparts.

%%%%% Reionization effect (vs Benitez+20)
In this context, it is important to note the seminal contributions from earlier studies on the effect of modeled reionization on star formation in halos.
For example, \cite{Benitez2015} and \cite{Benitez2020} showed that halos just above the cooling threshold can retain gas in equilibrium with the dark matter potential and UV background.
Those halos exhibit significantly delayed star formation histories, depending on their mass growth trajectory.
Recent models have also reported that the halo occupation fraction depends sensitively on reionization timing \citep{Kravtsov2022} and on the availability of molecular cooling in early-forming halos \citep{Nadler2025}.
Our study aligns with those findings, suggesting that the role of reionization is crucial in determining whether subhalos can host galaxies.
Beyond this, we also emphasize {\em the importance of the subhalo’s birthplace and early mass assembly history well before reionization.}

These conclusions connect with prior research, and we acknowledge that the key claims of this paper—(1) that reionization plays a crucial role in the “missing satellites” problem and (2) that a subhalo’s birth environment and mass accretion history determine whether it remains dark—are not entirely novel, even if approached from a different angle. 
The depth of our analysis, while significant, could benefit from statistically robust results across a larger sample of MWAs, allowing for a more detailed exploration of how self-shielding and gas properties influence the outcome from diverse environments.

Despite these connections with previous literature, this study presents some methodological differences and new perspectives.
For example, our simulations include more detailed modeling of stellar feedback compared to previous works, which could explain some discrepancies in the relative abundances of satellite galaxies.
However, the limited resolution of our simulations—particularly the absence of radiative transfer modeling for ionizing photons—prevents a more accurate treatment of self-shielding and high-density regions.
This issue, along with the neglect of local stellar radiation, suggests that further exploration of the self-shielding mechanisms and their impact on star formation would strengthen the findings.

Furthermore, while previous studies have primarily focused on isolated dwarf galaxies or a limited number of Milky Way-like simulations \citep{Munshi2017, Applebaum2021}, our approach has been applied to a larger sample of halos with similar masses, allowing us to evaluate object-to-object variations and better understand the role of the local environment alongside reionization in satellite survival within cosmological context.
A more detailed exploration of how these environmental factors, combined with the modeling choices (e.g., interstellar medium properties and star formation), influence the results would be a beneficial next step for future work.

% Limitation: resolution and SF modelling
Let us discuss the possible shortcomings of the present study.
One of the main constraints is the resolution of the simulations, which, although sufficient to resolve the internal structure of the most massive subhalos, may underestimate the effect of feedback in smaller galaxies.
Additionally, the adopted star formation model might not fully capture the stochastic variations in star formation within marginal-mass subhalos, suggesting that the results should be interpreted with caution in this regime.
However, these limitations do not invalidate the general conclusions but rather highlight areas where future research could improve our understanding of the dynamical evolution that affects the occupation fraction of subhalos and the faint-end of luminous satellites.

%%% Future expectations
Given these results, several directions for future studies emerge.
First, extending the analysis to higher-resolution simulations would be useful in order to achieve a better characterization of the population of lower-mass satellites.
Furthermore, a direct comparison with observational data, such as those from the Dark Energy Survey 5-yr SN analysis (DES-SN5YR, \citealt{Vincenzi2024}) or the Large Synoptic Survey Telescope (LSST, \citealt{Ivezic2019}) surveys, could provide even tighter constraints on feedback models and satellite survival.
Finally, future studies could explore the role of non-thermal processes, such as cosmic ray heating or the influence of the galactic magnetic field, which might have a non-negligible impact on the dynamics of subhalos and their ability to host visible galaxies.

We also acknowledge that our results depend on the specifics of the subgrid physics used in the simulation.
For example, our simulations lack full radiative transfer, which is crucial for tracing the local ionization and, in turn, gas temperature.
Recent studies based on radiative hydrodynamic simulations have indeed suggested that the reionization suppresses the gas mass growth and subsequent star formation in low-mass halos \citep{Onorbe2015, Ocvirk2020, Rey2020}.
Specifically, direct computation of ionizing photons could modify our results.
\citet{Katz2020} suggested that the hydrogen neutral fraction is affected by the large-scale structure (i.e., filaments), and \citet{Bhagwat2024} showed that a different scheme of supernova feedback, incorporating the radiative transfer, can change the reionization distribution and timing, and also the final stellar mass.
Using zoom-in simulation, \citet{Zier2025} compared the uniform UV background and the radiation field extracted from the parent full-box model and found the difference in ionization is much more severe in low-mass halos.
Moreover, given that low-mass halos dominantly contribute to the ionizing photon budget \citep{Lewis2020}, the ionization state of halos likely depends not only on the large-scale structure but also on the local neighborhood, adding complexity to the picture.
It would be exciting if such simulations could reach close to the local universe and cover a large cosmological volume in the near future.
Considering that \NH\ consumed nearly 100 million CPU hours of computing time, it is not feasible to run a similarly high-resolution simulation with radiative transfer.
However, if what we found in our study is largely true, that is, if the fate of starred and starless subhalos is determined so early, future studies could extend this work by exploring the epoch of reionization more thoroughly and monitoring galaxies during the first few gigayears.

%%%%%%%%%%%%%%%%%%%%%%%%%%%%%%%%%%%%%%%%%%%%%%%%%%%%%%%%%%%%%%%%%%%%%%%%%%%
%  _______  _______  __    _  _______  ___      __   __  ______   _______ 
% |       ||       ||  |  | ||       ||   |    |  | |  ||      | |       |
% |       ||   _   ||   |_| ||       ||   |    |  | |  ||  _    ||    ___|
% |       ||  | |  ||       ||       ||   |    |  |_|  || | |   ||   |___ 
% |      _||  |_|  ||  _    ||      _||   |___ |       || |_|   ||    ___|
% |     |_ |       || | |   ||     |_ |       ||       ||       ||   |___ 
% |_______||_______||_|  |__||_______||_______||_______||______| |_______|
%%%%%%%%%%%%%%%%%%%%%%%%%%%%%%%%%%%%%%%%%%%%%%%%%%%%%%%%%%%%%%%%%%%%%%%%%%%
\section{Conclusions}
\label{sec_conclusion}

Recent hydrodynamical simulations have shown that the use of updated baryonic physics suppresses the number of luminous satellite galaxies, going in the direction of solving the classical missing satellite problem.
In this work, we further investigate the origin of the observed dichotomy between luminous and dark subhalos using the \NH\ and \NHII\ simulations.
The high-resolution simulations indeed roughly reproduce the number of satellite galaxies in the Local Group.
We further explored the detailed origin of ``starless'' subhalos.
Following the dark halo mass function, less massive halos are more abundant.
In our simulations, starless subhalos are substantially less massive than starred ones.
We selected a sample of subhalos in a narrow mass range that contains both starred and starless subhalos and attempted to pin down what makes them starred or starless.

The key difference lies in their birthplace \textit{and} UV background heating probably associated with reionization.
When small subhalos are born in less gas-attracting regions, the gas within them struggles to condense.
Their fate is sealed by reionization, as only starred subhalos can maintain dense gas, self-shield, and continue to form stars.
The two most widely suspected candidate processes, e.g., supernova feedback and the orbital-motion associated environmental effects, were found to be irrelevant, primarily because starless subhalos never experienced in-situ star formation in the first place.
Moreover, these processes remove gas but not stars, and thus cannot transform star-forming subhalos into starless ones.
In conclusion, starless subhalos are born to be starless, not made.

This study contributes to the existing literature by providing new insights into the relationship between subhalo mass and satellite galaxy survival, with direct implications for the missing satellites problem and low-mass satellite galaxy formation modeling.
The results obtained confirm the trends observed in the works of \cite{Sawala2016}, \cite{Simpson2018}, and \cite{Benitez2019, Benitez2020}, while also suggesting new perspectives that warrant further investigation.
Hence, our results should trigger future research aimed at improving our understanding of the interplay between baryonic feedback, subhalo dynamical evolution, and star formation in Milky Way satellites.

%%%%%%%%%%%%%%%%%%%%%%%%%%%%%%%%%%%%%%%%%%%%%%%%%%%%%%%%%%%%%%%%%%%%%%%%%%%%%%%%%%%%%%%%%%%%%%%%%%%%%%%%%%%%%%%%%%%%%%%%%%

\section*{Acknowledgements}
We are particularly grateful to the referee for pointing us to numerous previous studies that were relevant to our investigation.
% NH (from Dubois2021, Yi2024)
This work was granted access to the HPC resources of CINES under the allocations c2016047637, A0020407637, and A0070402192 by Genci, KSC-2017-G2-0003, KSC-2020-CRE-0055, and KSC-2020-CRE-0280 by KISTI, and as a ``Grand Challenge'' project granted by GENCI on the AMD Rome extension of the Joliot Curie supercomputer at TGCC.
% NH2 (from Yi2024)
%S.J. has performed most of the analysis and writing.
%Y.D., S.K.Y., and S.H. ran the \NH\ and \NHII\ simulations.
%J.R. and Taysun Kimm implemented the chemical evolution in the \NHII\ simulation.
%E.C., K.K., S.P., and C.P. have contributed to the discussion of the results.
% KREONET (from Dubois2021)
The large data transfer was supported by KREONET, which is managed and operated by KISTI.
% SKY (from Rhee2024) 중견, 중점
S.K.Y. acknowledges support from the Korean National Research Foundation (RS-2025-00514475 and RS-2022-NR070872).
% Ema 창의
E.C. acknowledges support from the Korean National Research Foundation (RS-2023-00241934)
% Jinsu
J.R. was supported by the KASI-Yonsei Postdoctoral Fellowship and was supported by the Korea Astronomy and Space Science Institute under the R\&D program (Project No. 2025-1-831-02), supervised by the Korea AeroSpace Administration. 
This work was partially supported by the Institut de physique des Deux Infinis of Sorbonne Université and by the ANR grant ANR-19-CE31-0017 of the French Agence Nationale de la Recherche.

% SKIRT
We thank Maarten Baes for providing useful information for the radiative transfer code, SKIRT, which was used for generating mock images of our simulated galaxies.

%%%%%%%%%%%%%%%%%%%%%%%%%%%%%%%%%%%%%%%%%%%%%%%%%%%%%%%%%%%%%%%%%%%%%%%%
%  _______  _______  _______  _______  __    _  ______   ___   __   __ 
% |   _   ||       ||       ||       ||  |  | ||      | |   | |  |_|  |
% |  |_|  ||    _  ||    _  ||    ___||   |_| ||  _    ||   | |       |
% |       ||   |_| ||   |_| ||   |___ |       || | |   ||   | |       |
% |       ||    ___||    ___||    ___||  _    || |_|   ||   |  |     | 
% |   _   ||   |    |   |    |   |___ | | |   ||       ||   | |   _   |
% |__| |__||___|    |___|    |_______||_|  |__||______| |___| |__| |__|
%%%%%%%%%%%%%%%%%%%%%%%%%%%%%%%%%%%%%%%%%%%%%%%%%%%%%%%%%%%%%%%%%%%%%%%%
\appendix
\restartappendixnumbering

%%%%%%%%%%%%%%%%%%%%%%%%%%%%%%%%%%%%%%%%%%%%%%%%%%%%%%%%%%%%%%%%%%%%%%%%%%%
%%%%% 	[Merger Tree]
%%%%%%%%%%%%%%%%%%%%%%%%%%%%%%%%%%%%%%%%%%%%%%%%%%%%%%%%%%%%%%%%%%%%%%%%%%%
\section{Merger tree}\label{sec_method_tree}
High resolutions alone do not warrant accurate detection and classification of halos.
Building a merger tree for small halos, especially inside another halo, is not trivial because even very sophisticated merger tree algorithms cannot connect halos if they are not detected in the first place.
Besides, when a halo temporarily disappears or gets distorted due to a close encounter or merger, it may be linked to an incorrect progenitor.

Our main results, discussed in Section~\ref{sec_results}, track the very early stage of the halo evolution, so we need to be particularly careful with the merger tree.
% Ptree
First, we utilize the in-house merger tree, {\em PhantomTree} (Han, S., {\em priv. comm.}), based on the particle membership.
We generate a directed graph structure by treating halos as \textit{nodes} and connections as \textit{edges}.
For each halo (node), we find all descendant candidate halos where the member particles are at the next four snapshots, and then we calculate the score based on the fraction of the received and sent particles:
\begin{equation}
    S_{i,j} = \frac{n(i \cap j)}{n(j)}\times\frac{n(i \cap j)}{n(i)},
\end{equation}
where $S_{i,j}$ is the score between the target halo $i$ and the descendant candidate halo $j$, $n(i \cap j)$ is the number of shared member particles between them, $n(j)$ is the number of the member particles of the descendant candidate halo, and $n(i)$ is the number of the member particles of the target halo.
We directly connect two halos if they are mutually linked.
Otherwise, we connect them to the highest-scoring candidate node, which would be the merger or fragmentation cases.

% Rebuild with stable members
Second, we rebuild the merger tree using ``stable'' member particles that are defined when their occurrence exceeds the threshold in the main branch of the tree.
This occurrence threshold is initially set to the length of the branch but is iteratively reduced until stable member particles constitute more than 10\% of the number of members at the last snapshot.
To ensure more reliable connections in the tree, we re-identify the main progenitor, which has the most stable particles in each halo.
Based on these stable members, we also re-calculate the \textit{give} and \textit{take} scores:
\begin{equation}
    S_{{\rm give}, i} = \frac{N_i}{N},\
    S_{{\rm take}, i} = \frac{N_i}{n(i)},
\end{equation}
where $S_{\rm give, i}$ is the ``give'' score of the target halo $i$, $S_{\rm take, i}$ is the ``take'' score of the target halo $i$, $N_i$ is the number of stable member particles of the target halo $i$, $N$ is the total number of the stable member particles in the main branch, and $n(i)$ is the number of the particles of the target halo.
We discard the halo if $S_{\rm take}$ is less than 5\% or the $S_{\rm give} \times S_{\rm take}$ is less than 10\%.

We further exclude nodes with masses that deviate significantly from the local and global trends, as described below.
% Mass cut
We calculate the mean and standard deviation of the mass using the previous 100 nodes and the next 100 nodes in the main branch.
If the mass of the node is more than $4\,\sigma$ from the mean or 1,000 times the last mass, we tag it as a locally outlying node.
Also, if the mass of the node is more than $3\,\sigma$ from the median
% \ema{(Ema: formally speaking, $\sigma$ is not defined for a median. How about some confidence level, like percentiles?)}
of the main branch, we tag it as a globally outlying node.
We exclude both locally and globally outlying nodes from the main branch.

% Position cut
We then check the position of each node in the main branch.
By taking advantage of the dense cadence of the stored snapshots, we can estimate the expected position at the next snapshot.
If the node is located beyond the expected position by more than the sum of the virial radii of the two adjacent nodes, we further investigate.
For those nodes, we compare the raw positions of nodes and the polynomial fitted positions.
If the offset between the two positions is larger than the standard deviation of the residuals and exceeds 10 ckpc, we exclude the node from the main branch.
This process results in the reliable merger tree used in this work.

% Extend to initial "no halo" snapshots
We also extend the merger tree to the first snapshot where no halo is detected by tracking the position of stable particles, which is used in Figure~\ref{fig08_accretion} and \ref{fig09_gasevol}.
We acknowledge that this method has inherent uncertainties and cannot entirely avoid incorrect detections or broken tree issues, but repeated checks and strict criteria help minimize unwanted artifacts.

%%%%%%%%%%%%%%%%%%%%%%%%%%%%%%%%%%%%%%%%%%%%%%%%%%%%%%%%%%%%%%%%%%%%%%%%%%%
%%%%% 	[Validity of subhalo classification]
%%%%%%%%%%%%%%%%%%%%%%%%%%%%%%%%%%%%%%%%%%%%%%%%%%%%%%%%%%%%%%%%%%%%%%%%%%%
\section{Validity of subhalo classification}\label{sec_validity}

%=================================================================%
%          [Figure A1]
%=================================================================%
\begin{figure*}[htb!]
    \centering
    \includegraphics[width=0.95\textwidth]{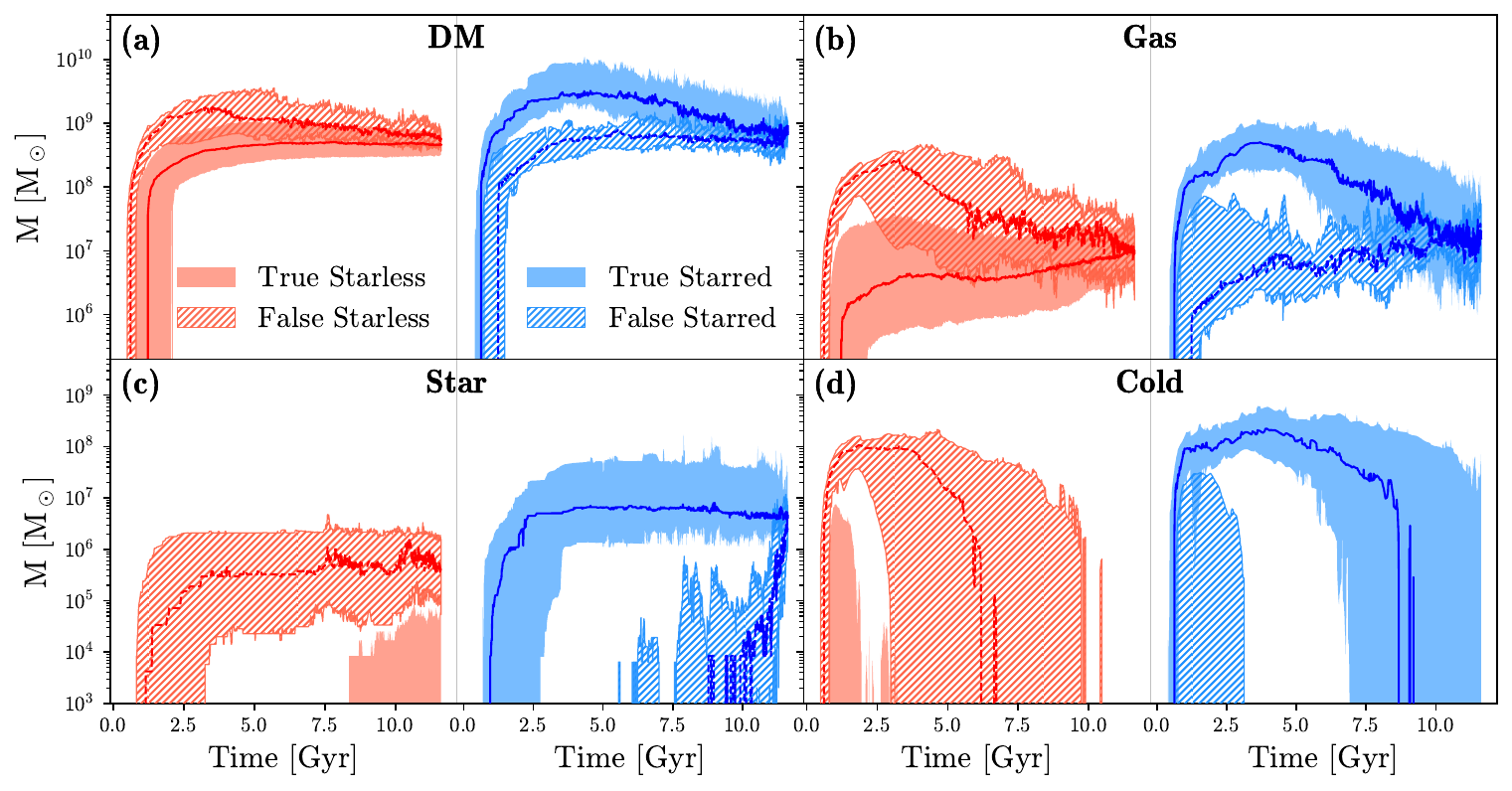}
    \caption{
        Comparisons of mass evolution of four populations with similar final DM masses.
        Starless subhalos are shown in red, and starred subhalos are shown in blue.
        Solid lines and shades represent the median and 16$^{\rm th}$ to 84$^{\rm th}$ percentiles of the \textit{true} populations.
        Dashed lines and hatched areas are those of the \textit{false} populations.
        % From panels (a) to (d), each panel
        Panels (a)–(d) show the mass evolution of DM, gas, stars, and cold gas, respectively.
        }
    \label{figA1_truefalse}
\end{figure*}
%=================================================================%
%=================================================================%

We introduce the classification of subhalos into ``starred'' and ``starless'' in Section~\ref{sec_method_classify} based on matching galaxies and halos.
However, this kind of automated method could lead to some misidentification.
Therefore, we further checked using their merger tree and star formation history to identify and correct false classifications.
We found \textit{false} starred subhalos that contain galaxies inside but show no star formation during their evolution.
These subhalos contain interloper stars from larger neighbors, and thus, they are numerically misidentified as starred subhalos.
We also identify \textit{false} starless subhalos that experienced significant star formation during their history.
This case is when subhalos do not have enough stars to be identified as galaxies.
We reclassify \textit{false} starred subhalos as starless subhalos and \textit{false} starless subhalos as starred subhalos.

To demonstrate the reliability of our classification, we show the mass evolution of the four populations, (true, false)$ \times$(starred, starless), in Figure~\ref{figA1_truefalse}.
We show the median and 16$^{\rm th}$ to 84$^{\rm th}$ percentiles of mass evolution for starred (blue) and starless (red) subhalos, distinguishing between \textit{true} (solid lines and shaded areas) and \textit{false} (dashed lines and hatched areas) populations.
It is clearly visible that the \textit{true} populations have distinct mass evolution trends from the \textit{false} populations.
For instance, in the stellar mass evolution (panel (c)) of the \textit{false} starless subhalos, there is significant star formation at the early epoch.
However, their stellar mass barely touches the minimum required to be detected as a galaxy, leading to their classification as starless subhalos.
Also, the \textit{false} starred subhalos have a lot of stars at the last snapshot, but we can infer that those are the interloper stars in the group systems.
The overall trends of the \textit{false} starless closely resemble those of the \textit{true} starred, and vice versa, which is the reason why we modified the original classification.

One might be concerned about the possibility that subhalos capture the field stars or stars from neighboring systems.
\cite{Penarrubia2024}, using analytic and N-body methods, demonstrated that subhalos in a galactic environment do capture field stars and could keep them permanently under certain conditions.
However, our subhalo samples have various infall epochs to the main system, and before that, we find that capturing events rarely happen.
Truly, most cases of \textit{false} starless and \textit{true} starred subhalos have in-situ star formation, which contributes most of their final stellar components.

%%%%%%%%%%%%%%%%%%%%%%%%%%%%%%%%%%%%%%%%%%%%%%%%%%%%%%%%%%%%%%%%%%%%%%%%%%%
%%%%% 	[Supernova energy and rates]
%%%%%%%%%%%%%%%%%%%%%%%%%%%%%%%%%%%%%%%%%%%%%%%%%%%%%%%%%%%%%%%%%%%%%%%%%%%
\section{Supernova energy and rates}\label{sec_snpower}
In the main text, we analyze the impact of supernova feedback with supernova power, $P_{\rm SN}$.
In this section, we want to describe the definition of it.
As prescribed in the subgrid model of supernova feedback (see the detailed process in Section 2.4 of \citep{Dubois2021}), a young stellar particle releases energy and mass during a supernova explosion, following the Chabrier initial mass function (IMF) \citep{Chabrier2003}.
At each timestep, we calculate the total supernova energy, $\rm{E_{SN}}$, by summing the energy released from young stars within the boundary of each subhalo.
The supernova energy rate, $\epsilon_{\rm SN}$, at each timestep ($t_i$) is calculated as:
\begin{equation}
    \epsilon_{\rm SN}(t_i) = \frac{{\rm E_{SN}} (t_i)}{\Delta t_i},
\end{equation}
where $\Delta t_i$ is the time interval between the $i^{\text{th}}$ and $(i-1)^{\text{th}}$ snapshots.
To quantify both the amplitude and duration of supernova events, we define the ``supernova power,'' $P_{\rm SN}$, by averaging these energy rates over timesteps where $\rm{E_{SN}} (t_i) > 0$.
This physically indicates an expected value of supernova energy for a given time interval if supernovae occur.
For instance, a high value of $P_{\rm SN}$ indicates that a burst of supernovae occurred over a short period of time.

%%%%%%%%%%%%%%%%%%%%%%%%%%%%%%%%%%%%%%%%%%%%%%%%%%%%%%%%%%%%%%%%%%%%%%%%%%%
%%%%% 	[Radiative cooling and heating]
%%%%%%%%%%%%%%%%%%%%%%%%%%%%%%%%%%%%%%%%%%%%%%%%%%%%%%%%%%%%%%%%%%%%%%%%%%%
\section{Radiative cooling and heating}\label{sec_cooling}

As we have demonstrated the suppression of radiative cooling due to UV background heating in starless subhalos in the main text, we hereby provide a detailed description of the radiative cooling and heating prescription.
Since radiative transfer is computationally expensive and not applied in \NH\ and \NHII, our code computes the density evolution of six species (e$^-$, HI, HII, HeI, HeII, and HeIII) and then constructs the cooling and heating tables, taking into account the bremsstrahlung, ionization, recombination, dielectric recombination, line cooling, radiative heating, and Compton cooling and heating (model described in \citep{Courty2004}).
We reconstruct the cooling and heating rates based on the density and temperature of each cell using the stored tables:
\begin{equation}
    \Lambda_{\rm net} = \Lambda_{\rm cool} + (Z/Z_\odot)\Lambda_{\rm metal} - \Gamma_{\rm heat} + (\Lambda_{\rm Comp}-\Gamma_{\rm Comp})/n_{\rm H},
\end{equation}
\begin{equation}
    \Lambda_{\rm net}^\prime(=\frac{d\Lambda_{\rm net}}{dT}) = \Lambda_{\rm cool}^\prime + (Z/Z_\odot)\Lambda_{\rm metal}^\prime - \Gamma_{\rm heat}^\prime + (\Lambda_{\rm Comp}^\prime-\Gamma_{\rm Comp}^\prime)/n_{\rm H},
\end{equation}
where $\Lambda_{\rm net}$ and $\Lambda_{\rm net}^\prime$ are the net cooling rate and its derivative.
$\Lambda_{\rm cool}$ and $\Lambda_{\rm metal}$ are the cooling rates of the gas and the metal, $\Gamma_{\rm heat}$ is the heating rate, $\Lambda_{\rm Comp}$ and $\Gamma_{\rm Comp}$ are the Compton cooling and heating rates, and $n_{\rm H}$ is the number density of hydrogen.

%=================================================================%
%          [Figure A2]
%=================================================================%
\begin{figure*}[htb!]
    \centering
    \setcounter{figure}{0}
    \includegraphics[width=1.\textwidth]{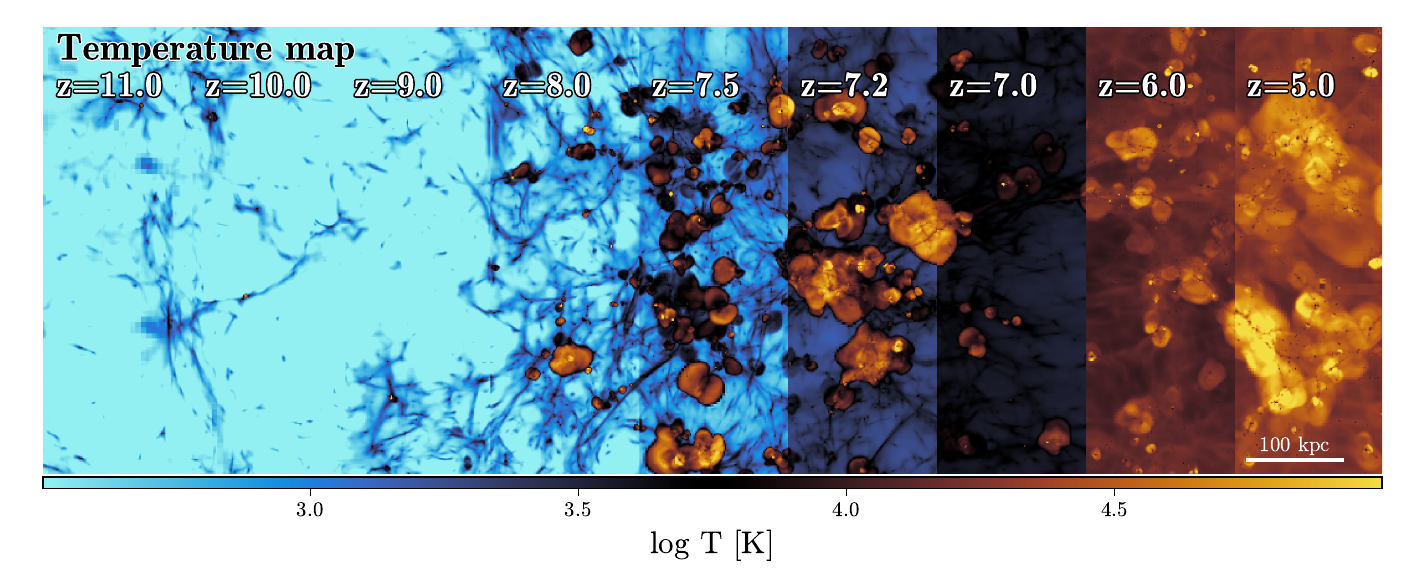}
    \caption{
        Sequential temperature map of the \NH\ simulation, as it visually demonstrates the stage of reionization.
        From left to right, each slice indicates a different redshift from 11 to 5.
        Redder colors represent higher temperatures.
        }
    \label{figA2_tempmap}
\end{figure*}
%=================================================================%
%=================================================================%

The uniform UV background heating after $z_{\rm reion}=10$ is implicitly incorporated into the radiative cooling and heating tables \citep{Rosdahl2012}, rather than through direct radiative transfer.
The self-shielding effect is included as a boosting factor as follows:
\begin{equation}
    n_{\rm H, corr} = {n_{\rm H} / e^{-n_{\rm H}/(0.01{\rm H\,cm^{-3}})}}
\end{equation}
where $n_{\rm H, corr}$ is the corrected number density of hydrogen that is practically used to compute the cooling and heating rates, and self-shielding threshold of $0.01{\rm H\,cm^{-3}}$ is motivated by result of \cite{Rosdahl2012}.
In Figure~\ref{figA2_tempmap}, we display the sequential temperature map for high redshift snapshots.
Although the uniform UV background turns on at $z=10$, it is prominent after $z\sim7$.

%=================================================================%
%          [Figure A3]
%=================================================================%
\begin{figure*}[htb!]
    \centering
    \includegraphics[width=1.\textwidth]{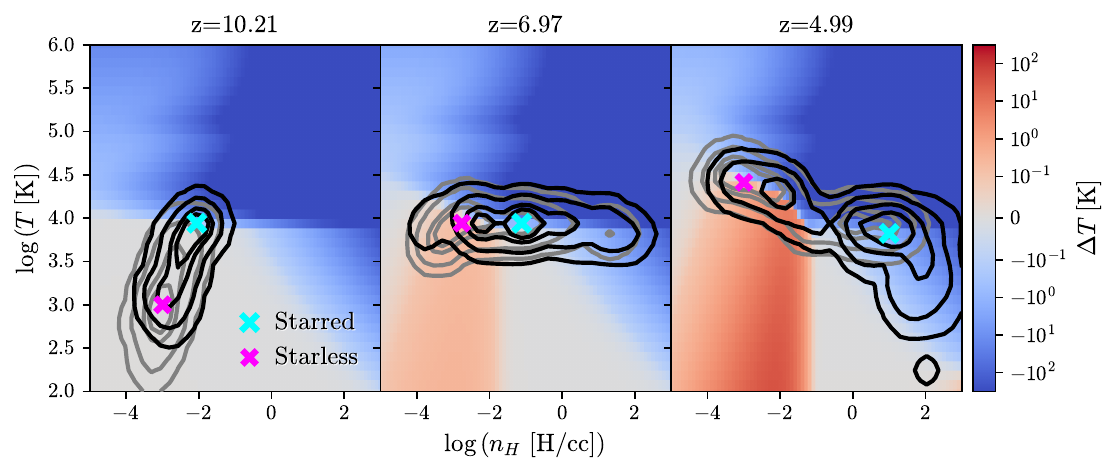}
    \caption{
        The temperature changes within a single computational time step, following the radiative cooling and heating prescriptions.
        The metallicity is assumed to be 0.001 times the solar metallicity.
        The three panels for the three different epochs marked at the top.
        We color-coded the net temperature change across the parameter space (cell density and temperature).
        Contours represent the 0.5, 1, 2, and 3 $\sigma$ distributions of dense cells of starless (grey) and starred (black) subhalos.
        The dense cells are defined as the top 25 densest within the virial radius.
        Markers represent peak values in gas phase space.
        }
    \label{figA3_cooling}
\end{figure*}
%=================================================================%
%=================================================================%

Figure~\ref{figA3_cooling} shows the reconstructed change of the temperature based on the above procedures, {\em excluding} adiabatic processes.
We compute the net change in temperature for different gas densities and temperatures.
Bluer colors represent net cooling, and redder colors represent net heating.
The three panels show three different epochs around reionization.
The heating zones are most sensitive to UV background heating.
A sharp transition from cooling to heating occurs around a density of \hcc{0.01}, corresponding to the self-shielding threshold.
This effect becomes more pronounced around redshift 7 when the epoch of reionization is almost completed.
The overall distribution of dense cells in starless (grey and magenta) and starred (black and cyan) subhalos is also shown with contours and markers.
Around $z=7$, the divergence is noticeable, and the starless subhalos still remain in the heating zones, while starred subhalos have already entered the cooling zones.
%%%%%%%%%%%%%%%%%%%%%%%%%%%%%%%%%%%%%%%%%%%%%%%%%%%%%%%%%%%%%%%%%%%%%%%%%%%%%%%%%%%%%%%%%%%%%%%%%%%%%%%%%%%%%%%%%%%%%%%%%%

\bibliography{reference}{}
\bibliographystyle{aasjournal}

\end{document}